\documentclass[]{article}

\usepackage{booktabs}
\usepackage{multirow}
\usepackage{amssymb} 
\usepackage{amsbsy}
\usepackage{mathtools}
\usepackage{microtype}
\usepackage[table]{xcolor}
\usepackage{enumerate}
\usepackage{authblk}
\usepackage[utf8]{inputenc}
\usepackage[margin=1in]{geometry}
\usepackage[round]{natbib}
\usepackage{float}
\usepackage{url}

\usepackage{hhline}

\providecommand{\abs}[1]{\lvert#1\rvert}

\makeatletter
  \newcommand\scripty{\@setfontsize\scriptsize{6pt}{8}}
\makeatother
\makeatletter
  \newcommand\scriptx{\@setfontsize\scriptsize{7pt}{8}}
\makeatother
\makeatletter
  \newcommand\scriptz{\@setfontsize\scriptsize{5pt}{6}}
\makeatother
\usepackage{dsfont}

\usepackage{xr}

\makeatletter
\newcommand*{\addFileDependency}[1]{
  \typeout{(#1)}
  \@addtofilelist{#1}
  \IfFileExists{#1}{}{\typeout{No file #1.}}
}
\makeatother

\definecolor{green!30}{RGB}{178, 255, 178}
\definecolor{red!30}{RGB}{255, 178, 178}
\definecolor{blue!30}{RGB}{178, 178, 255}

\begin{document}





\title{Pairwise versus multiple global network alignment} 

\author[ ]{Vipin Vijayan\hspace{-0.75ex}}
\author[$\ddag$]{Shawn Gu}
\author[$\ddag$]{Eric Krebs}
\author[ ]{Lei Meng\hspace{-0.75ex}}
\author[*]{Tijana Milenkovi\'{c}}

\affil[ ]{Department of Computer Science and Engineering}
\affil[ ]{Eck Institute for Global Health}
\affil[ ]{Interdisciplinary Center for Network Science and Applications}
\affil[ ]{University of Notre Dame}
\affil[$\ddag$]{These authors equally contributed to this work}
\affil[*]{To whom correspondence should be addressed (email: tmilenko@nd.edu)}
\date{}
\maketitle

\abstract{
Biological network alignment (NA) aims to identify similar regions between molecular networks of different species. NA can be local or global. Just as the recent trend in the  NA field, we also focus on global NA, which can be pairwise (PNA) and multiple (MNA). PNA produces aligned node pairs between two networks. MNA produces aligned node clusters between more than two networks. Recently, the focus has shifted from PNA to MNA, because MNA captures conserved regions between more networks than PNA (and MNA is thus considered to be more insightful), though at higher computational complexity. The issue is that, due to the different outputs of PNA and MNA, a PNA method is only compared to other PNA methods, and an MNA method is only compared to other MNA methods.  Comparison of PNA against MNA must be done to evaluate whether MNA's higher complexity is justified by its higher accuracy. We introduce a framework that allows for this. We evaluate eight prominent PNA and MNA methods, on synthetic and real-world biological networks, using topological and functional alignment quality measures. We compare PNA against MNA in both a pairwise (native to PNA) and multiple (native to MNA) manner. PNA is expected to perform better under the pairwise evaluation framework. Indeed this is what we find.  MNA is expected to perform better under the multiple evaluation framework. Shockingly, we find this not to always hold; PNA is often better than MNA in this framework, depending on the choice of evaluation test. 
}
\maketitle

\section{Introduction}\label{sec:intro}

\subsection{Motivation and background}

Networks can be used to model complex real-world systems in many  domains, including computational biology. A popular type of biological networks are protein interaction networks (PINs). While PIN data are available for multiple species \citep{BioGRID}, the functions of many proteins in many species remain unknown \citep{NAtransfer1,NAtransfer2}.  Network alignment (NA) compares networks to find a node mapping that conserves similar regions between the networks.  Then, analogous to genomic sequence alignment, NA can be used to predict protein functions by transferring functional knowledge from a well-studied species to a poorly-studied species between the species' conserved (aligned) PIN regions \citep{FaisalAging,ConeReview,ElmsallatiReview,LocalVsGlobal,GuzziNA}. While we focus on biological NA of PINs, NA can be used for many applications \citep{FiftyYears}, including computer vision \citep{NApaperDuchenne}, online social networks \citep{COSNET}, and ontology matching \citep{NApaperBayatiJ}.

NA is related to the subgraph isomorphism, or subgraph matching, problem. This problem asks to find a node mapping such that one network is an exact subgraph of another network. NA is a more general problem in that it asks to find a node mapping that best ``fits'' one network into another network, even if the first network is not an exact subgraph of the second. A widely used measure that quantifies this ``fit'' is the amount of conserved (aligned) edges, i.e., the size of the common conserved subgraph between the aligned networks. Since maximizing edge conservation is NP-hard \citep{MIGRAAL}, heuristic methods are needed for NA.

Like genomic sequence alignment, NA can be {\em local} or {\em global} \citep{LocalVsGlobal,GuzziNA}.  Initial research was on local NA, which searches for \emph{small highly conserved} regions across the compared networks, irrespective of the overall similarity between the networks; the conserved network regions can, but are not required to, overlap. More recent efforts have focused on global NA, which searches for a node mapping that maximizes overall similarity of the compared networks and thus results in \emph{large} but \emph{suboptimally conserved} network regions. Each of local NA and global NA has its (dis)advantages \citep{LocalVsGlobal,IGLOO,GuzziNA}. Because in the recent years global NA has received more attention than local NA, in this paper we also focus on global NA, and henceforth, we refer to global NA as NA.

Also, and importantly for our study, NA methods can be {\em pairwise} or {\em multiple} \citep{ConeReview,GuzziNA}.  While pairwise NA (PNA) aligns two networks at once, multiple NA (MNA) can align more than two networks at once. Since MNA can capture conserved network regions between multiple networks, it is hypothesized that MNA may lead to deeper biological insights compared to PNA.  On the other hand, MNA is computationally harder than PNA since the complexity of the NA problem can increase exponentially with the number of considered networks. However, this hypothesis has not been tested yet (for reasons described below). Because of this, and because both PNA and MNA have the same ultimate goal, which is to transfer knowledge from well- to poorly-studied species, we argue that they need to be compared in order to determine which category of methods produce superior alignments.

\begin{figure}
\centering
\includegraphics[width=0.8\linewidth]{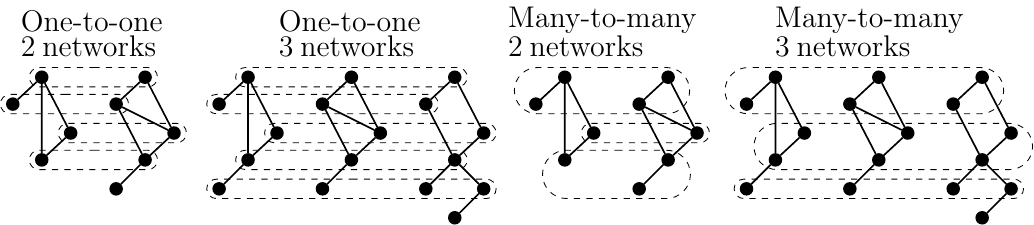}
\caption{Illustration of different alignment types: one-to-one alignments between two networks and between more than two (in this case three) networks, and many-to-many alignments between two networks and between more than two (in this case three) networks.  For a one-to-one alignment, a node in one network is mapped to at most one node in another network, and a node cannot be mapped to another node in the same network.  For a many-to-many alignment, a node in one network can be mapped to multiple nodes in another network, and a node can also be mapped to other nodes in the same network.}
\label{fig:topo}
\end{figure}

Since typical PNA and MNA methods produce alignments of different types (Fig. \ref{fig:topo}), it has been difficult to compare them. Namely, when aligning two networks, PNA typically produces a {\em one-to-one} node mapping between the two networks, which results in aligned node {\em pairs}.  When aligning more than two networks, MNA produces a node mapping across the multiple networks, which results in aligned node {\em clusters}.  If an aligned cluster contains more than one node from a single network, then it is a {\em many-to-many} alignment. If each of the aligned clusters contains at most one node per network, then it is a {\em one-to-one} alignment. Typical MNA methods produce many-to-many alignments, and they are called many-to-many MNA methods. MNA methods that produce one-to-one alignments are called one-to-one MNA methods. MNA methods can also be trivially used to align pairs of networks, which results in aligned node clusters for many-to-many MNA methods and in aligned node pairs for one-to-one MNA methods.

There is sometimes confusion in the literature that one-to-one alignments are automatically global (i.e., outputted by global NA methods), and that many-to-many alignments are automatically local (outputted by local NA methods). However, this is not necessarily the case. First, one-to-one alignments can result in only small regions aligned to each other (clearly without any nodes overlapping), meaning that they are local one-to-one alignments. Second, many-to-many alignments can result in aligned node clusters covering nodes from all analyzed networks, meaning that they are global, many-to-many alignments. In other words, in our opinion, ``local'' and ``global'' describe how much of the networks' nodes are covered by (i.e., are a part of) the given alignment, and  not on whether the nodes are aligned in  one-to-one or many-to-many fashion. It is important to note that most of the recent one-to-one methods will not actually produce local alignments, because they require all nodes of the smaller networks to be mapped to nodes of the larger networks, automatically leading to global (one-to-one, or even more formally, injective) alignments. However, this is an algorithmic design choice of many existing methods rather than a requirement of any and every one-to-one method. As  discussed above, we focus on global NA, considering both one-to-one and many-to-many methods.


Again, because PNA and MNA generally produce alignments of different types (aligned node pairs versus aligned node clusters, respectively), alignment quality measures designed for alignments of one type do not necessarily work for alignments of the other type. Also, alignment quality measures designed for alignments of two networks do not necessarily work for alignments of more than two networks. Due to this difficulty, when a new PNA or MNA method is proposed, it is only compared against other NA methods from the same category. However, since both PNA and MNA have the same goal of across-species knowledge transfer, we argue that there is a need to compare them. This is especially true because early evidence suggests that aligning each pair of considered networks via PNA and then combining the pairwise alignments into a multiple alignment spanning all of the networks can be superior to directly aligning all networks via MNA \citep{SMAL}.



\subsection{Our contributions}\label{sect:contributions}

\begin{figure}
\centering
\includegraphics[width=\linewidth]{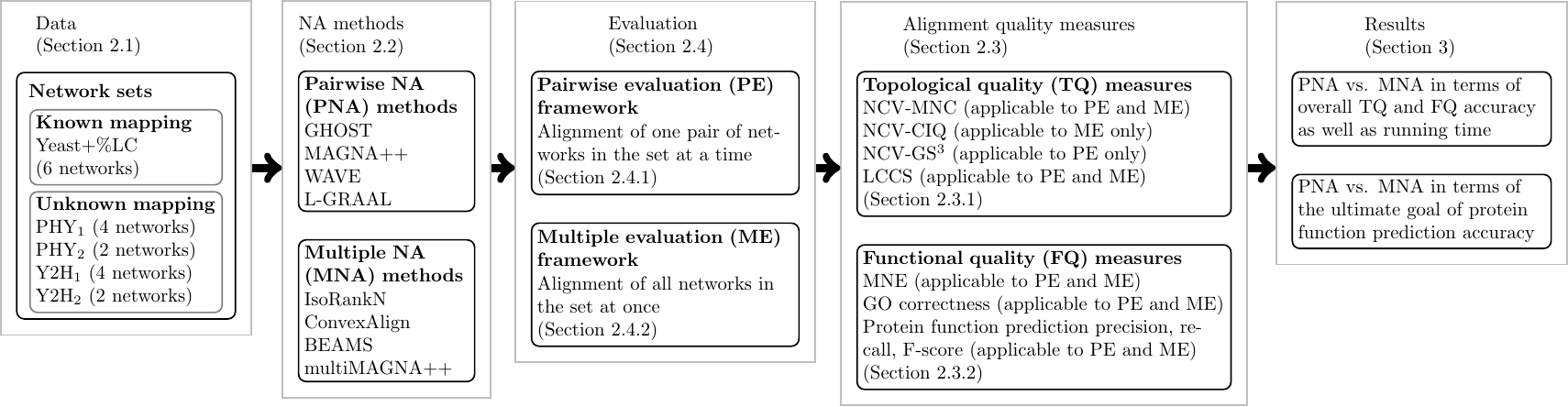}
\caption{Overview of our PNA versus MNA evaluation framework. We   evaluate prominent existing PNA and MNA methods (Section   \ref{sec:aligners}) on synthetic and real-world biological networks   (Section \ref{sec:dataset}), in terms of both topological (TQ) and   functional (FQ) alignment quality (Section \ref{sec:quality}) and   running time. First, we evaluate in a pairwise manner (native to PNA) under the pairwise evaluation (PE) framework (Section \ref{sec:pef}).  Here, for the given network set, we (i) trivially apply PNA to network pairs (which we denote as PE-P-P alignments), (ii) trivially apply MNA to network pairs (PE-M-P), and (iii) apply MNA to the whole network set (all networks at once) and break the resulting multiple alignment into pairwise alignments (PE-M-M). We also evaluate in a multiple manner (native to MNA) under the multiple evaluation (ME) framework (Section \ref{sec:mef}). Here, we (i) apply PNA to network pairs and combine the resulting pairwise alignments into a multiple alignment (ME-P-P), (ii) apply MNA to network pairs and combine the resulting pairwise alignments into a multiple alignment (ME-M-P), and (iii) trivially apply MNA to the whole network set (ME-M-M).  Finally, we compare PNA and MNA in   terms of protein function prediction accuracy, which is the ultimate   goal of biological NA.  }
\label{fig:flow}
\end{figure}

Thus, we propose an evaluation framework for a fair comparison of PNA
and MNA, and we use it to comprehensively compare the two, in both a
pairwise (native to PNA) and multiple (native to MNA) manner (Fig.
\ref{fig:flow}).
%

We evaluate prominent PNA and MNA methods that were published by the beginning of our study, were publicly available, and had user-friendly implementations.  This includes four PNA methods (GHOST \citep{GHOST}, MAGNA++ \citep{MAGNA++}, WAVE \citep{WAVE}, and L-GRAAL \citep{LGRAAL}), and four MNA methods (IsoRankN \citep{IsoRankN}, BEAMS \citep{BEAMS}, multiMAGNA++ \citep{multiMAGNA++}, \textcolor{black}{and ConvexAlign} \citep{ConvexAlign}). Most of these methods are recent and were thus already shown to be superior to many past methods, e.g., IsoRank \citep{IsoRank}, MI-GRAAL \citep{MIGRAAL}, GEDEVO \citep{GEDEVO}, and NETAL \citep{NETAL} PNA methods, plus GEDEVO-M \citep{GEDEVOM}, FUSE \citep{FUSE}, and SMETANA \citep{SMETANA} MNA methods. Note that newer NA methods have appeared since, such as SANA \citep{SANA}, ModuleAlign \citep{ModuleAlign},  SUMONA \citep{SUMONA}, and PrimAlign \citep{kalecky2018primalign}, which is why they were not included here. Importantly, we believe that their inclusion is not required. \textcolor{black}{This is because our goal is \textbf{not} to determine the best existing (PNA or MNA) method. Instead, it is to properly evaluate the whole category of prominent recent PNA methods against the whole category of equally prominent recent and thus fairly comparable MNA methods. While the best existing NA method would likely change with introduction of each new method (or possibly even a new  measure for evaluating alignment quality), the best category of NA approaches is less likely to change, unless there is a drastic shift in how the NA problem is approached and solved (or possibly even just how alignment quality is evaluated). And one of the purposes of our study is to determine if such a shift is needed.}

We evaluate the PNA and MNA methods on synthetic networks with known true node mapping (we know the underlying alignment that a perfect method should output) and real-world PINs of different species with unknown node mapping (we do not know which protein in one species corresponds to which protein in the other species).

We evaluate alignment quality using topological and functional alignment quality measures. An alignment is of good topological quality if it reconstructs well the underlying true node mapping (when known) and if it has many conserved edges (i.e., if it conserves a large common subgraph between the networks).  An alignment is of good functional quality if its aligned node pairs/clusters contain nodes with similar biological functions.

Section \ref{sec:methods} describes the data, alignment quality measures, and evaluation framework.  Section \ref{sec:results} describes our findings.

\section{Methods}
\label{sec:methods}

\subsection{Data}
\label{sec:dataset}

We use five network sets: one synthetic network set with known true node mapping, and four real-world network sets with unknown true node mapping. For each network, we use only its largest connected component.

\vspace{0.25em}\noindent{\bf Network set with known true node mapping.}  This synthetic network set, named Yeast+\%LC, contains a high-confidence {\em S. cerevisiae} (yeast) PIN with $1,004$ proteins and $8,323$ interactions \citep{YeastLC}, along with five lower-confidence yeast PINs constructed by adding 5\%, 10\%, 15\%, 20\%, or 25\% of lower-confidence interactions to the high-confidence PIN (Supplementary Table \ref{tab:nets}).  This network set has been used in many existing studies \citep{GRAAL,HGRAAL,MIGRAAL,GHOST,MAGNA,LocalVsGlobal,multiMAGNA++}. Since all networks have the same node set, we know the true node mapping. Hence, for this set, we can evaluate node correctness, i.e., how well the given NA method reconstructs the true node mapping (Section \ref{sec:topoquality}).

\vspace{0.25em}\noindent{\bf Network sets with unknown true node mapping.}  The four real-world network sets with unknown node mapping are named PHY$_1$, PHY$_2$, Y2H$_1$, and Y2H$_2$.  Each contains PINs of four species, {\em S. cerevisiae} (yeast), {\em D. melanogaster} (fly), {\em C.   elegans} (worm), and {\em H. sapiens} (human). The PIN data, obtained from BioGRID \citep{BioGRID}, have been used in recent studies \citep{LocalVsGlobal,multiMAGNA++}.  For each species, four PINs are created that contain the following protein interaction types and confidence levels: all physical interactions supported by at least one publication (PHY$_1$) or at least two publications (PHY$_2$), as well as only yeast two-hybrid physical interactions supported by at least one publication (Y2H$_1$) or at least two publications (Y2H$_2$) (Supplementary Table \ref{tab:nets}). Just as was done in the existing studies, we also remove the fly and worm networks from the PHY$_2$ and Y2H$_2$ network sets, because these networks are too small and sparse (53-331 nodes and 33-260 edges), resulting in the PHY$_2$ and Y2H$_2$ network sets containing only two networks each.  The four network sets have unknown true node mapping, and thus we cannot evaluate node correctness. However, we use alternative measures of alignment quality that are based on Gene Ontology annotations (Section \ref{sec:funcquality}).

\vspace{0.25em}\noindent{\bf Gene Ontology (GO) annotations.} For alignment quality measures (Section \ref{sec:quality}) that rely on GO annotations of proteins \citep{GOdata}, we use experimentally obtained GO annotations from the GO database from January 2016.

\vspace{0.25em}\noindent {\bf Protein sequences.}  When NA methods use protein sequence information to produce an alignment (Section \ref{sec:aligners}), we use BLAST protein sequence similarities as captured by E-values \citep{BLAST}. The sequence data were acquired from the NCBI website (https://www.ncbi.nlm.nih.gov/).


\subsection{NA methods that we evaluate}
\label{sec:aligners}

We study GHOST, MAGNA++, WAVE, and L-GRAAL PNA methods, and IsoRankN, BEAMS,  multiMAGNA++, \textcolor{black}{and ConvexAlign} MNA methods.

\vspace{0.25em}\noindent{\bf PNA methods.} Most NA methods are {\em two-stage} aligners: first, they calculate the similarities (based on network topology and, optionally, protein sequences) between nodes of the compared networks, and second, they use an alignment strategy to find high scoring alignments with respect to the total similarity over all aligned nodes.  GHOST is a two-stage PNA method (Supplementary Section \ref{supplsec:aligners}).  An issue with two-stage methods is that while they find high scoring alignments with respect to total node similarity (a.k.a. node conservation), they do not account for the amount of conserved edges during the alignment construction process. But the quality of an alignment is often measured in terms of edge conservation.  To address this, MAGNA++ directly optimizes both edge and node conservation {\it while} the alignment is constructed (Supplementary Section \ref{supplsec:aligners}).  MAGNA++ is a {\em search-based} (rather than a two-stage) PNA method.  Search-based aligners can directly optimize edge conservation or any other alignment quality measure. WAVE and L-GRAAL were proposed as two-stage (rather than search-based) PNA methods that, just as MAGNA++, optimize both node and (weighted) edge conservation (Supplementary Section \ref{supplsec:aligners}).


\vspace{0.25em}\noindent{\bf MNA methods.} IsoRankN,  BEAMS, \textcolor{black}{and ConvexAlign}  are two-stage MNA methods. IsoRankN \textcolor{black}{optimizes node conservation. BEAMS and ConvexAlign optimize} both node and edge conservation (Supplementary Section \ref{supplsec:aligners}). On the other hand, like MAGNA++, multiMAGNA++ is a search-based method that optimizes both edge and node conservation. \textcolor{black}{ IsoRankN and BEAMS produce many-to-many alignments. ConvexAlign and multiMAGNA++ produce one-to-one alignments.}

\vspace{0.25em}\noindent{\bf Aligning using network topology only versus using both   topology and protein sequences.} In our analysis, for each method, we study the effect on output quality when (i) using only network topology while constructing alignments (T alignments) versus (ii) using both network topology and protein sequence information while constructing alignments (T+S alignments).  For T alignments, we set method parameters to ignore any sequence information. All methods except BEAMS can produce T alignments and all methods can produce T+S alignments.  For T+S alignments, we set method parameters to include sequence information. Supplementary Table \ref{tab:params} shows the specific parameters that we use, and Supplementary Section \ref{supplsec:aligners} justifies our parameter choices.


\subsection{Alignment quality measures}
\label{sec:quality}

Typical PNA methods produce alignments comprising node pairs and typical MNA methods produce alignments comprising node clusters.  We introduce the term {\em aligned node group} to describe either an aligned node pair or an aligned node cluster.  With this, we can represent a pairwise or multiple alignment as a set of aligned node groups. For formal definitions, see Supplementary Section \ref{supplsec:quality}.


\subsubsection{Topological quality (TQ) measures}
\label{sec:topoquality}
 
A good NA method should produce aligned node groups that have internal consistency with respect to protein labels. If we know the true node mapping between the networks, we can let the labels be node names.  We consider measures that rely on node names to be capturing topological quality (TQ) of an alignment.  If we do not know the true node mapping, we let the labels be nodes' (i.e., proteins') GO terms. We consider measures that rely on GO terms to be capturing functional quality (FQ) of an alignment; we discuss such measures in Section \ref{sec:funcquality}. We measure internal consistency of aligned protein groups in a \emph{pairwise} alignment via precision, recall, and F-score of node correctness (P-NC, R-NC, and F-NC, respectively); these measures, introduced by \citet{LocalVsGlobal}, work for both one-to-one and many-to-many pairwise alignments (Supplementary Section \ref{supplsec:topoquality}). We do this in a \emph{multiple} alignment via adjusted multiple node correctness (NCV-MNC); this measure, introduced by \citet{multiMAGNA++}, works for both one-to-one and many-to-many multiple alignments (Supplementary Section \ref{supplsec:topoquality}).

Also, a good NA method should find a large amount of common network structure, i.e., produce high edge conservation. We measure edge conservation in a \emph{pairwise} alignment via adjusted generalized S$^3$ (NCV-GS$^3$); this measure, introduced by \citet{LocalVsGlobal}, works for both one-to-one and many-to-many pairwise alignments (Supplementary Section \ref{supplsec:topoquality}). We do this in a \emph{multiple} alignment via adjusted cluster interaction quality (NCV-CIQ); this measure, introduced by \citet{multiMAGNA++}, works for both one-to-one and many-to-many multiple alignments (Supplementary Section \ref{supplsec:topoquality}).

Finally, for a good NA method, conserved edges should form large and dense (as opposed to small or isolated) conserved regions.  We capture the notion of large and connected conserved network regions (for \emph{both} pairwise and multiple alignments) via largest common connected subgraph (LCCS). This measure, recently extended from PNA \citep{MAGNA} to MNA \citep{multiMAGNA++}, works for both one-to-one and many-to-many alignments, and for both pairwise and multiple alignments (Supplementary Section \ref{supplsec:topoquality}).


\subsubsection{Functional quality (FQ) measures}
\label{sec:funcquality}

Per Section \ref{sec:topoquality}, a good alignment should have internally consistent aligned node groups.  Instead of protein names as in Section \ref{sec:topoquality}, in this section we use GO terms as protein labels to measure internal consistency. Having aligned node groups that are internally consistent with respect to GO terms is important for protein function prediction.

We measure internal node group consistency with respect to GO terms in two ways. First, we do so via mean normalized entropy (MNE); this measure, introduced by \citet{IsoRankN} (also, see \citet{multiMAGNA++} for formal definition), works for both one-to-one and many-to-many alignments, and for both pairwise and multiple alignments (Supplementary Section \ref{supplsec:funcquality}). Second, we do so via  an alternative popular measure, GO correctness (GC); this measure, recently extended from PNA \citep{GRAAL} to MNA \citep{multiMAGNA++}, works for both one-to-one and many-to-many alignments, and for both pairwise and multiple alignments (Supplementary Section \ref{supplsec:funcquality}).

In addition to measuring internal node group consistency, we directly measure the accuracy of protein function prediction.  That is, we first use a protein function prediction approach (Section \ref{sec:prediction}) to predict protein-GO term associations, and then we compare the predicted associations to known protein-GO term associations to see how accurate the predicted associations are. We do so via precision, recall, and F-score measures (P-PF, R-PF, and F-PF, respectively); these measures work for both one-to-one and many-to-many alignments, and for both pairwise and multiple alignments (Supplementary Section \ref{supplsec:funcquality}).


\subsubsection{Protein function prediction approaches}
\label{sec:prediction}

Here, we discuss how we predict protein-GO term associations from the given alignment. We use a different protein function prediction approach for each alignment type. Therefore, below, first, we discuss an existing approach that we use to predict protein GO-term associations from pairwise alignments (approach 1).  Second, we discuss an existing approach that we use to predict these associations from multiple alignments (approach 2). Third, since the existing approach for multiple alignments (approach 2) is very different from the existing approach for pairwise alignments (approach 1), to make comparison between pairwise and multiple alignments (i.e., between PNA and MNA) more fair, we extend approach 1 for pairwise alignments into a new approach for multiple alignments (approach 3). As we show in Section \ref{sec:newpreds}, our new approach 3 \textcolor{black}{in general} improves upon the existing approach 2. So, we propose approach 3 as a new superior strategy for predicting protein-GO term associations from multiple alignments, which is another contribution of our study.

\vspace{0.25em}\noindent{\bf Approach 1. Existing protein function prediction for pairwise alignments.}  Here, we predict protein GO-terms associations using a multi-step process proposed by \citet{LocalVsGlobal}.  For each protein $v$ in the alignment that has at least one annotated GO term, and for each GO term $g$, first, we hide $v$'s true GO term(s). Second, we determine if the alignment is statistically significant with respect to $g$, i.e., if the number of aligned node pairs in which the aligned proteins share GO term $g$ is significantly high ($p$-value below 0.05 according to the hypergeometric test; see \citep{LocalVsGlobal} for details). Repeating this process for all nodes and GO terms results in set $X$ of predicted protein-GO term associations.

\vspace{0.25em}\noindent{\bf Approach 2. Existing protein function prediction for multiple alignments.}  Here, we predict protein GO-term associations using the approach of \citet{FaisalAging}, as follows. For each protein $v$ in the alignment that has at least one annotated GO term, and for each GO term $g$, first, we hide the protein's true GO term(s). Second, given that $v$ belongs to aligned node group $C$, we measure the enrichment of $C$ in $g$ using the hypergeometric test. If $C$ is significantly enriched in $g$ ($p$-value below 0.05; see \citep{multiMAGNA++} for details), then we predict $v$ to be associated with $g$. Repeating this process for all nodes and GO terms results in set $X$ of predicted protein-GO term associations.

\vspace{0.25em}\noindent{\bf Approach 3. New protein function prediction for multiple alignments.}  Here, we introduce a new approach to predict protein GO-term associations from a multiple alignment.  First, for each node group $C_i$ in the alignment, $C_i$ is converted into a   set of all possible $\abs{C_i} \choose 2$ node pairs in the   group. The union of all resulting node pairs over all groups $C_i$   forms the set $F$ of all aligned node pairs.  Second, for each   protein $v$ in the alignment that has at least one annotated GO   term, and for each GO term $g$, we hide $v$'s true GO term(s).   Third, we determine if the alignment is statistically significant   with respect to $g$, i.e., if the number of aligned node pairs $F$   in which the aligned proteins share GO term $g$ is significantly   high ($p$-value below 0.05 according to the hypergeometric test; see   Supplementary Section \ref{sec:suppl:ppf2} for details).  Repeating   this process for all nodes and GO terms results in a set of   predicted protein-GO term associations.  Our proposed approach 3 is   identical to approach 1 except for its first step of converting a   multiple alignment into a set of aligned node pairs.


\subsubsection{Statistical significance of alignment quality scores}
\label{sec:significance}

Since PNA and MNA methods result in different output types (as they produce alignments that differ in the number and sizes of aligned node groups for the same networks), to allow for as fair as possible comparison of the different NA methods, we do the following. For each NA method, each pair/set of aligned networks, and each alignment quality measure, we compute the statistical significance (i.e., $p$-value) of the given alignment quality score. Then, we take the significance of each alignment quality score into consideration when comparing the NA methods (as explained in Section \ref{sec:ranking}). We compute the $p$-value of a quality score of an alignment as described in Supplementary Section \ref{supplsec:significance}.


\subsection{Evaluation framework}
\label{sec:evaluation}

Given a network set, to fairly compare PNA and MNA, we compare the NA methods when aligning all possible pairs of networks in the set (pairwise evaluation framework, Section \ref{sec:pef}), as well as when aligning all networks in the set at once (multiple evaluation framework, Section \ref{sec:mef}). PNA is expected to perform better under the pairwise evaluation framework (which is native to PNA), and MNA is expected to perform better under the multiple evaluation framework (which it is native to MNA).


\subsubsection{Pairwise evaluation (PE) framework}
\label{sec:pef}

In the {\em PE framework}, given a network set, we compare NA methods using pairwise alignments of all possible pairs of networks in the set. Due to the various ways that a pairwise alignment of two networks can be created using PNA or MNA methods, we categorize the pairwise alignments into the following three categories. Specifically, we: 
\begin{itemize} 
\item Apply PNA to all possible network pairs, denoting the resulting alignments as the PE-P-P alignment category. \textcolor{black}{Here, since all PNA methods are one-to-one,  their pairwise alignments will be one-to-one.} 
\item Apply MNA to all possible network pairs, denoting the resulting alignments as the PE-M-P alignment category. \textcolor{black}{Here, if an MNA method is many-to-many, then its pairwise alignments will also be many-to-many. Otherwise, they will be one-to-one.} 
\item Apply MNA to the whole network set and break the resulting multiple alignment into all possible pairwise alignments (Fig. \ref{fig:pemm}(a)), denoting the resulting pairwise alignments as the PE-M-M alignment category. \textcolor{black}{Again, for a one-to-one or many-to-many MNA method,  its pairwise alignments will also be one-to-one or many-to-many, respectively.} 
\end{itemize} 
In the PE framework, we align all pairs of networks within each of the five analyzed network sets (Yeast+\%LC, PHY$_1$, PHY$_2$, Y2H$_1$, and Y2H$_2$; Section \ref{sec:dataset}). We evaluate using all alignment quality measures for pairwise alignments, namely F-NC, NCV-GS$^3$, and LCCS TQ measures as well as MNE, GC, and F-PF FQ measures (Section \ref{sec:quality}).



\subsubsection{Multiple evaluation (ME) framework}
\label{sec:mef}

\begin{figure}[p]
\center
{\bf(a)}\includegraphics[width=0.6\linewidth]{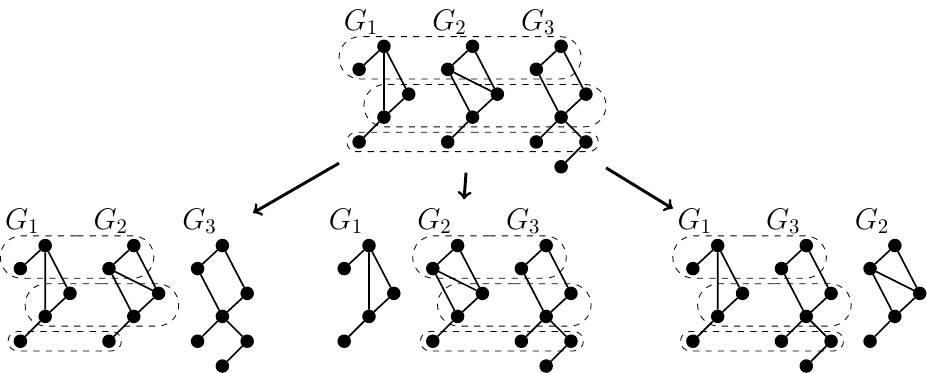} \vspace{2.5mm}  \\
{\bf(b)}\includegraphics[width=0.6\linewidth]{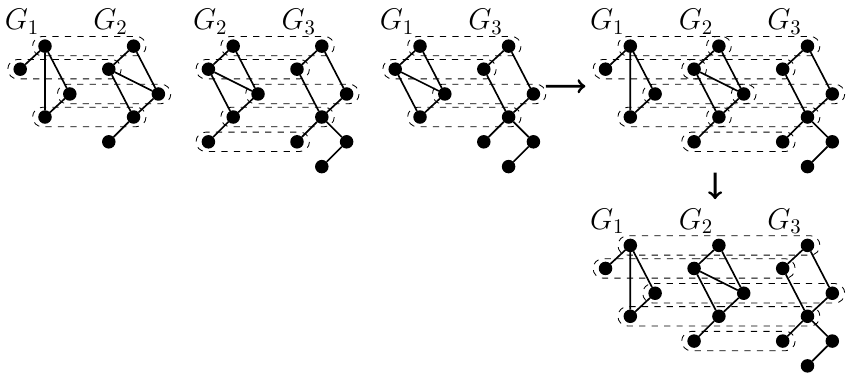} \vspace{2.5mm} \\
{\bf(c)}\includegraphics[width=0.6\linewidth]{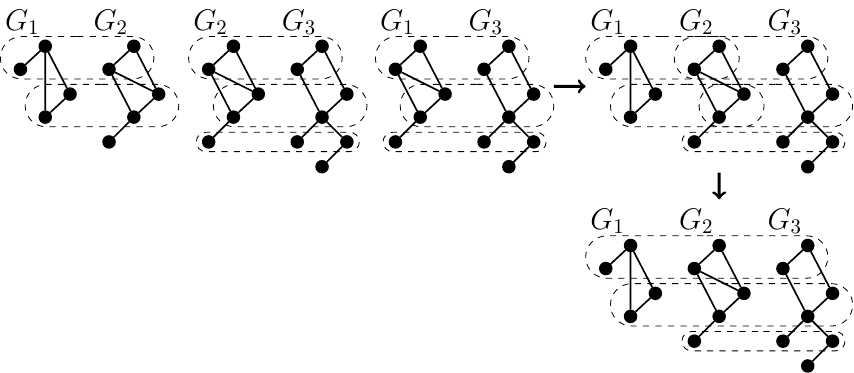}
\caption{Illustration of how we convert from one NA output type to another. Given a network set consisting of three networks ($G_1$, $G_2$, and $G_3$), we convert: {\bf(a)} a multiple alignment to pairwise alignments, {\bf(b)} one-to-one pairwise alignments to a multiple alignment, and {\bf(c)} many-to-many pairwise alignments to a multiple alignment.  (a) Given a multiple alignment spanning all three networks, we create a pairwise alignment for every pair of networks (in our case, three pairs) as follows: for the given two networks, we remove every node from the multiple alignment that is not a part of the two networks, which results in a pairwise alignment of the two networks.  (b,c) Given pairwise alignments of all networks pairs in the set (here, three pairs of networks, ($G_1$,$G_2$), ($G_2$,$G_3$), and ($G_1$,$G_3$)), produced by either (b) PNA or (c) MNA, we combine the pairwise alignments into a multiple alignment as follows.  First, we select a ``scaffold'' network (here, $G_2$).  Second, we create a set of node groups consisting of the pairwise alignments between the scaffold network and the other networks (here, ($G_1$,$G_2$) and ($G_2$,$G_3$)).  Third, we merge node groups that have at least one node in common.  This procedure yields a multiple alignment of all networks in the set.  }
\label{fig:pemm}
\end{figure}

 In the {\em ME framework}, given a network set, we compare NA methods using the resulting multiple alignments of the set. Due to the various ways that a multiple alignment of a network set can be created, we categorize the multiple alignments in the following three categories. Specifically, we: 
 \begin{itemize} 
 \item Apply PNA to all possible network pairs and combine the resulting pairwise alignments into a multiple alignment that spans all networks in the set using a variation of a method introduced by \citet{SMAL} (Fig. \ref{fig:pemm}(b)-(c) and Supplementary Section \ref{supplsec:evaluation}), denoting the resulting alignments as the ME-P-P alignment category. \textcolor{black}{Here, even though all PNA methods are one-to-one, their pairwise-combined-to-multiple alignments will be many-to-many.} 
 \item Apply MNA to all possible network pairs and combine the resulting pairwise alignments into a multiple alignment that spans all networks in the set using the same variation of the method introduced by \citet{SMAL} as above (Fig. \ref{fig:pemm}(b)-(c) and Supplementary Section \ref{supplsec:evaluation}), denoting the resulting alignments as the ME-M-P alignment category. \textcolor{black}{Here, independent of whether an MNA method is one-to-one or many-to-many, its pairwise-combined-to-multiple alignments will be many-to-many.}  
 \item Apply MNA to the whole network set to align all   networks at once, denoting the resulting alignments as the ME-M-M   category. \textcolor{black}{Here, if an MNA method is one-to-one, its direct multiple alignments will also be one-to-one. Otherwise, they will be  many-to-many.} 
 \end{itemize} 
 In the ME framework, we align each of the analyzed network sets that has more than two networks (Yeast\%+LC, PHY$_1$, and Y2H$_1$; Section \ref{sec:dataset}). We evaluate using all alignment quality measures for multiple alignments, namely NCV-MNC, NCV-CIQ, and LCCS TQ measures as well as MNE, GC, and F-PF FQ measures (Section \ref{sec:quality}).


\subsubsection{Comparing the performance of NA methods}
\label{sec:ranking}

Given a network pair/set and an alignment quality measure (i.e., in a given {\em evaluation test}), we compare two NA methods as follows. Let $x$ and $y$ be the methods' respective alignment quality scores. If both $x$ and $y$ are significant ($p$-values below 0.001; Section \ref{sec:significance}) and are within 1\% of each other ($\frac{\abs{x-y}}{(x+y)/2} < 0.01$), then the two methods are {\em   tied}. They are also {\em tied} if both $x$ and $y$ are non-significant.  If both $x$ and $y$ are significant and not tied, then the method with the best score is superior. If $x$ is significant and $y$ is not, then the method with score $x$ is superior, and vice versa.

Given $k$ network pairs/sets and $l$ alignment quality measures, i.e., given $k \times l$ evaluation tests, for each evaluation test, we rank all methods from the best one to the worst one, as follows.  Given the methods' alignment quality scores, for methods with non-significant scores, we rank the methods last.  For methods with significant scores, we perform the following procedure. If a given method has the best alignment quality score, then we give it rank 1 (as the 1$^{st}$ best method).  We give the next best performing method rank 2, and so on.  If a given method is tied with the next best performing method, then we rank both methods with the superior (i.e., lower) rank. The subsequent methods are ranked as if the previous methods were not tied. For example, if methods $a$ and $b$ are tied, they are both given rank 1, and if method $c$ is not tied with method $a$ or method $b$, then method $c$ is given rank 3).  We call this resulting rank for a given evaluation test an {\em evaluation test rank}.  We calculate the \emph{overall ranking} of an NA method by taking the mean of its ranks over all $k \times l$ evaluation tests. To evaluate whether the overall rankings of two methods are significantly different from each other, we apply the one-tailed Wilcoxon signed-rank test on the $k \times l$ evaluation test ranks of the two methods.



\section{Results and discussion}
\label{sec:results}


In Section \ref{sec:tvsts}, we compare the quality of T alignments and T+S alignments. 
In Sections \ref{sec:exp2} and \ref{sec:exp1}, we compare PNA against MNA in the PE and ME framework, respectively, in terms of TQ and FQ accuracy as well as running time.
In Section \ref{sec:predsmain}, we compare PNA against MNA exclusively in terms protein function prediction accuracy, as the main goal of biological NA is to predict protein functions in one species from protein functions in another species, based on the species' network alignment.

\subsection{T versus T+S alignments}
\label{sec:tvsts}

\begin{figure}[ht!]
\centering
\includegraphics[width=0.7\linewidth]{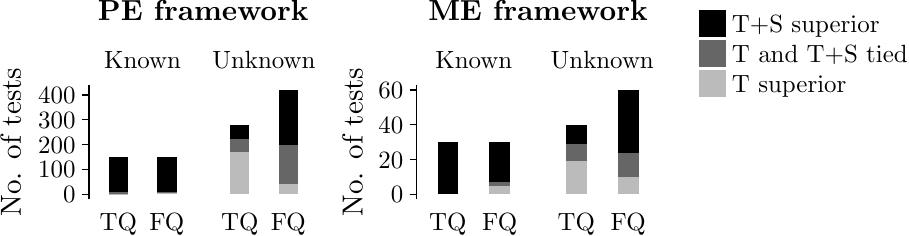}
\caption{Comparison of the quality of T alignments versus the   corresponding T+S   alignments, under each of the PE and ME frameworks.  Each bar shows the number of evaluation tests (out of all conducted tests, where a test is a combination of an NA method, a network pair/set, and an alignment quality measure) in which the T alignment is superior, the T+S alignment is superior, or the two alignments are tied (i.e., within 1\% of each other's accuracy). The evaluation tests are separated into network pairs/sets with known true node mapping and network pairs/sets with unknown true node mapping, and into TQ and FQ measures.}
\label{fig:t_vs_ts}
\end{figure}

Network topology alone can be used to find good alignments of PINs \citep{GRAAL}. But protein sequence information can be used to complement network topology in order to produce superior alignments \citep{Complementarity}.  Due to the complementarity of network topology and protein sequence information, we expect T+S alignments to have higher alignment quality than T alignments.  In fact, we verify this.  Namely, for each NA method, we compare the given method's T alignments to their corresponding T+S alignments, in terms of TQ and FQ measures, under the PE and ME frameworks (Fig. \ref{fig:t_vs_ts}).  We find the following.

For networks with known true node mapping, T+S alignments are superior to the corresponding T alignments in almost all cases. Note that as already recognized by \citet{multiMAGNA++}, for these networks, i.e., for the Yeast+\%LC network set, the superiority of T+S alignments over T alignments is not a surprising result. This is because this dataset contains networks that all have the same set of nodes. Consequently, it contains many inter-network pairs of nodes that are the same proteins.  Sequence similarities of such matching node pairs are higher than those of other non-matching node pairs. These matching inter-network node pairs can likely form aligned node groups that have very high intra-group sequence similarity due to the node pairs containing identical proteins. This could explain the superiority of T+S alignments over T alignments for the set of networks with known node mapping.  

Even for the sets of networks with unknown node mapping (PHY1, PHY2, Y2H1, Y2H2), whose networks contain different node sets, we still see that T+S alignments are overall superior to T alignments. Namely, \textcolor{black}{only in terms of TQ, T alignments are somewhat superior to T+S alignments, but T+S alignments are still superior to or tied with the corresponding T alignments in just under a half of all cases}.  In terms of FQ, T+S alignments are superior to or tied with the T alignments in almost all evaluation tests.

So, we conclude that T+S alignments are overall superior to T alignments. Because of this, because T+S alignments are more relevant in the computational biology domain, and because of space constraints,  henceforth, we mainly analyze T+S alignments. Importantly, our findings for T+S alignments also hold for T alignments (Supplementary Fig. \ref{fig:res_t}).


\textcolor{black}{Due to space constraints, for additional results on the similarity (overlap) of the alignments produced the different NA methods, which demonstrate that using protein sequence information overall yields alignment consistency between the different NA methods, see Supplementary Section \ref{sec:suppl_tvsts} and Supplementary  Figs. \ref{fig:expboth_dendro}--\ref{fig:exp1_dendro}. }

\subsection{Method comparison in the PE framework}
\label{sec:exp2}

We expect that under the PE framework, PNA will perform better than MNA. This is exactly what we observe. So, the most interesting and shocking results of our study do not originate from this section. Instead, they originate from Section \ref{sec:exp1} below, when comparing PNA and MNA in the ME framework.

Namely, in the PE framework, the overall ranking of the PNA methods (T+S alignments from the PE-P-P category) is generally better (lower) than the overall ranking of the MNA methods (T+S alignments from the PE-M-P and PE-M-M categories) (View I of Fig. \ref{fig:res}). An exception is multiMAGNA++'s alignments from the PE-M-P category (multiMAGNA++ directly applied to network pairs), whose overall ranking is \textcolor{black}{also} very good (low). This \textcolor{black}{could be} due to multiMAGNA++ being a one-to-one MNA method, which \textcolor{black}{might have} caused it to behave similarly \textcolor{black}{as PNA methods (all of which are also one-to-one)} when it is used to align only two networks. \textcolor{black}{This is further supported by the fact that the only other considered one-to-one MNA method, ConvexAlign, and specifically its PE-M-P version, is also ranked better (lower) than the remaining two many-to-many MNA methods, IsoRankN and BEAMS. Nonetheless, ConvexAlign still has worse (higher) ranking than any PNA method (View I of Fig. \ref{fig:res}).}

Next, we break down the results into those for networks with known versus unknown node mapping, and also, into those for TQ versus FQ measures (View II of Fig.  \ref{fig:res}).  
For networks with known mapping, we find that PNA performs better than MNA in terms of both TQ and FQ. For networks with unknown mapping, PNA performs better than MNA in terms of TQ, while in terms of FQ, \textcolor{black}{the situation is not as clear.}

\textcolor{black}{Namely, for networks with unknown mapping and FQ, as can be seen in View II of Fig.  \ref{fig:res}, MNA falls into the best  (lowest) ranks 1-4 in more of the evaluation tests than PNA.  This implies that MNA is better than PNA. However, at the same time, MNA also falls into the worst (highest) ranks 9-12 in more of the evaluation tests than PNA. This implies that MNA is worse than PNA. Because we are interested in comparing the whole category of the considered PNA approaches against the whole category of the considered MNA approaches (per our discussion in Section \ref{sect:contributions}), the above two results combined could be interpreted as MNA and PNA being comparable for networks with unknown mapping and FQ. On the other hand, for the same networks (with unknown mapping) and TQ, as well as for networks with known mapping and both TQ and FQ, PNA  falls into the best  ranks 1-4 in more of the evaluation tests than MNA, and at the same time, PNA falls into the worst ranks 9-12 in \emph{fewer} of the evaluation tests than MNA, which means that PNA is superior to MNA.}

\textcolor{black}{Another observation is as follows (Supplementary Tables \ref{tab:exp2_ranktable_ts_all_topo}-\ref{tab:exp2_ranktable_ts_unknown_all}). For evaluation tests in which PNA is clearly superior in terms of method rankings to MNA (again, with the exception of multiMAGNA++'s PE-M-P version), which are tests excluding networks with unknown mapping and FQ, the best-ranked PNA  method  (MAGNA++ or WAVE) is significantly superior to the best-ranked MNA method (multiMAGNA++'s PE-M-M version, followed by all other MNA methods that are all similarly ranked), with $p$-values below $1.8 \times 10^{-6}$. On the other hand, for tests where it is unclear which of PNA and MNA is better, which are tests involving networks with unknown mapping or FQ, the best-ranked MNA method (ConvexAlign's PE-M-P version) is only marginally  better than the best-ranked PNA method (MAGNA++), with $p$-values between 0.048 and 0.332. This justifies referring to PNA and MNA as comparable for networks with unknown mapping and FQ, and to PNA as being superior in all other cases.}

\textcolor{black}{Next, we want to comment on the two MNA methods that perform well in at least some evaluation tests in the PE (\emph{pairwise}) framework: multiMAGNA++ and ConvexAlign. Both of these methods produce one-to-one mappings, unlike the other two MNA methods, BEAMS and IsoRankN, which produce many-to-many mappings. Given that all PNA (\emph{pairwise}) methods are also one-to-one, it might not be surprising that the two one-to-one MNA methods also perform well in the PE framework. This could be because the existing  measures for \emph{pairwise} alignment accuracy favor one-to-one mappings. However, we believe that it is not just the one-to-one aspect of multiMAGNA++ and ConvexAlign that is relevant. First, while multiMAGNA++ performs reasonably well in all tests (networks with both known and unknown node mappings, and both TQ and FQ), ConvexAlign performs poorly for networks with known mapping or TQ but exceptionally well (marginally better than multiMAGNA++) for networks with unknown mapping and FQ. So, even though both methods are one-to-one, each has its unique (dis)advantages. Second, in Section \ref{sec:exp1}, which evaluates the methods in the ME (\emph{multiple}) framework, of the four MNA methods, it is again multiMAGNA++ and ConvexAlign that perform the best. This is despite the fact that the existing  measures for \emph{multiple} alignment accuracy do \emph{not} necessarily favor one-to-one mappings, and some (especially  FQ) actually \emph{favor} many-to-many mappings.}

\textcolor{black}{A likely reason why ConvexAlign  performs well only for networks with unknown node mapping and FQ is because its parameter values that were recommended and pre-set by its authors and that we use (Supplementary Section \ref{supplsec:aligners}) were determined via cross-validation, by optimizing FQ (GO term similarity of mapped nodes) in alignments of networks with unknown node mapping (PPI networks of mouse and human) \citep{ConvexAlign}. Hence, ConvexAlign is semi-supervised, i.e., pre-trained to achieve high FQ scores, which makes it biased compared to the other considered NA methods, all of which are unsupervised. }


\noindent{\bf Accuracy versus running time.}  The PNA methods are not only more accurate in general (as demonstrated above), but on average they are also at least somewhat if not much faster (View III of Fig. \ref{fig:res}).  In fact, no MNA method has both better running time and better ranking than any PNA method, while many PNA methods have both better running time and better ranking than every MNA method.


\subsection{Method comparison in the ME framework}
\label{sec:exp1}

We expect that under the ME framework, MNA will perform better than PNA.  Shockingly, we do not find this. Instead, our results reveal the opposite trends, which match those observed under the PE framework. So, the most interesting results of our study originate from this section. 

\begin{figure}
\centering
\includegraphics[width=1\linewidth]{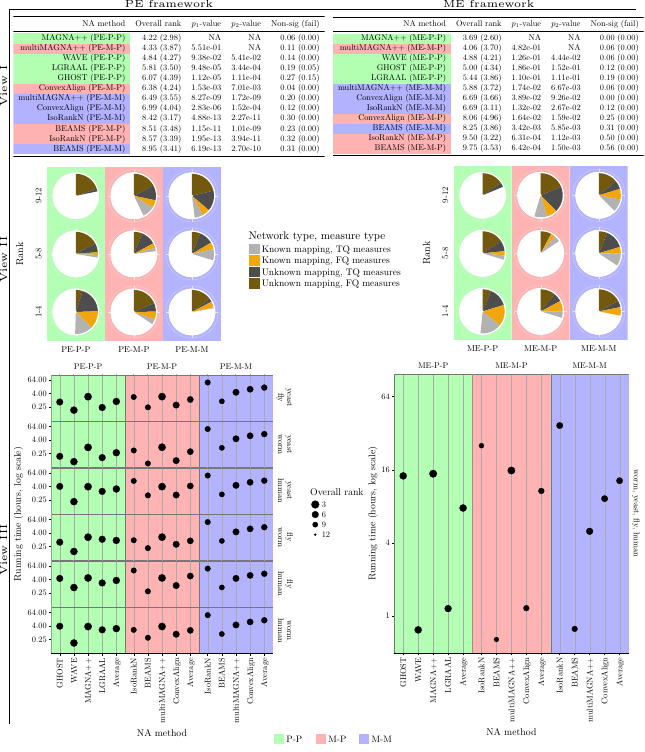}
\caption{Method comparison results for each of the PE and ME frameworks over all evaluation tests (where a test is a combination of an NA method, a network pair/set, and an alignment quality measure). The figure legend is continued on the next page.}
\label{fig:res}
\end{figure}

Namely, in the ME framework, the overall ranking of the PNA methods (T+S alignments from the ME-P-P category) is generally better (lower) than the overall ranking of the MNA methods' T+S alignments from the ME-M-M category, which in turn is generally better than the overall ranking of the MNA methods' T+S alignments from the ME-M-P category (View I of Fig. \ref{fig:res}).  Again, multiMAGNA++ is an exception: its alignments from the ME-M-P category (multiMAGNA++ first being applied to network pairs and then its pairwise alignments being combined into a multiple alignment) are ranked very good (low).

When we inspect the ranking of the methods in more detail (View II of Fig. \ref{fig:res}), again, we find similar trends as in the PE framework.  Namely, for networks with known mapping, we find that PNA performs better than MNA in terms of both TQ and FQ.  For networks with unknown mapping, PNA performs better than MNA in terms of TQ. In terms of FQ, \textcolor{black}{just as under the PE framework, MNA falls into the best (lowest) ranks in more of the evaluation tests than PNA, but at the same time, MNA also falls into the worst (highest) ranks in more of the evaluation tests than PNA. }

\textcolor{black}{Another result also applies to the ME framework: of the MNA methods, multiMAGNA++ and ConvexAlign perform better than BEAMS and IsoRankN, where multiMAGNA++ performs consistently well across all tests, and ConvexAlign performs extremely well only for networks with unknown node mapping and FQ (Supplementary Tables \ref{tab:exp1_ranktable_ts_all_topo}-\ref{tab:exp1_ranktable_ts_unknown_all}). }

\textcolor{black}{Notice that under the ME framework, the best (PNA or MNA) methods are all one-to-one. Because all considered PNA methods are one-to-one, one might suspect that PNA may be overall better than MNA in the ME framework not because of the ``pairwise'' part but simply because of the ``one-to-one'' part, possibly because one might suspect our evaluation measures in the ME framework to favor one-to-one methods. However, we argue that this is not the case, as follows. }

\textcolor{black}{First, if we could show that any existing one-to-one method performed worse than any existing many-to-many method in our ME framework, this would suffice to show that our ME framework does not favor one-to-one-methods. While for our considered methods it is the case that one-to-one (PNA or MNA) methods are superior to many-to-many (MNA) methods, this could be simply because the considered one-to-one methods are more recent and thus more powerful than the considered many-to-many methods. Indeed, when we add to our ME evaluation an older (and thus inferior) one-to-one MNA method, GEDEVO-M \citep{GEDEVOM}, we find that this one-to-one method is  outperformed by the considered many-to-many MNA methods (Supplementary Tables \ref{tab:exp1_gedevom_ts_all_topo}-\ref{tab:exp1_gedevom_ts_all_all}). If one-to-one methods had some advantage over many-to-many methods in our ME framework, this would not have happened. So, a method's performance in our ME framework does not seem to be directly related to it being one-to-one or many-to-many.}

\textcolor{black}{Second, by design, our evaluation measures do not favor one-to-one methods. Namely, recall that many of our  evaluation measures were proposed by studies that introduced or analyzed many-to-many NA methods (Section \ref{sec:quality}). An example is one of our considered FQ measures, mean normalized entropy (MNE), which originates from the IsoRankN study \citep{IsoRankN}, where IsoRankN is one of the considered many-to-many MNA methods. So, MNE is unlikely to favor one-to-one methods, as it was proposed in the many-to-many context. Actually, when we mirror the exact same MNE evaluation as in the IsoRankN study (see \cite{IsoRankN} for details) on the methods we consider here (rather than combine MNE with our other FQ measures as done so far in the paper), the considered one-to-one methods still perform well (i.e., the best of all considered one-to-one methods is still better than the best of all considered many-to-many methods) (Supplementary Tables \ref{tab:entropy_t}-\ref{tab:entropy_ts}). That is, even a measure designed explicitly for many-to-many alignments still ranks one-to-one-alignments better than many-to-many alignments. This additionally confirms that the overall superiority of the considered one-to-one (PNA or MNA) methods over the considered many-to-many (MNA) methods in the ME framework is likely because the one-to-one methods actually yield higher-quality alignments.}




\textcolor{black}{In summary, with these two findings in mind, it is more likely that the considered one-to-one methods perform better than the considered many-to-many methods in the ME framework because recent studies have focused on one-to-one alignments. Consequently, increased research in this area has likely led to better methodological advancements of one-to-one methods compared to many-to-many methods, explaining the one-to-one methods' superior performance.}

%


\noindent{\bf Accuracy versus running time.} When we compare the overall rankings of the NA methods to their running times (View III of Fig. \ref{fig:res}), again, we find similar trends as in the PE framework: the PNA methods are not only more accurate (as demonstrated above), but on average they are also faster. 

Since the PNA methods must align every pair of networks in order to produce a multiple alignment, and since this results in a quadratically increasing running time with respect to the number of networks $k$, we ask whether there is some value of $k$ at which PNA might become less efficient (i.e., slower) than MNA. \textcolor{black}{Due to space constraints, we present this discussion in Supplementary Section \ref{sect:suppl_ME_runtime} and Supplementary Table \ref{tab:complexity}.}

\pagebreak
\noindent Figure \ref{fig:res}: Method comparison results for each of the {\bf PE} and {\bf ME} frameworks over all evaluation tests (where a test is a combination of an NA method, a network pair/set, and an alignment quality measure, for T+S alignments).  By NA method, here, we mean the combination of a PNA or MNA method and the alignment category (Section \ref{sec:evaluation}). Namely, there are 12 NA methods in the PE framework (four PNA methods associated with the PE-P-P category and four MNA methods associated with each of the PE-M-M and PE-M-P categories) and 12 NA methods in the ME framework (four PNA methods associated with the ME-P-P category and four MNA methods associated with each of the ME-M-M and ME-M-P categories). The alignment categories are color coded.  {\bf View I.} Overall ranking of the NA methods. The ``Overall rank'' column shows the rank of each method averaged over all evaluation tests, along with the corresponding standard deviation (in brackets). Since there are 12 methods in a given framework, the possible ranks range from 1 to 12. The lower the rank, the better the given method.  The ``$p_1$-value'' column shows the statistical significance of the difference between the ranking of each method and the 1$^{st}$ best ranked method.  The ``$p_2$-value'' column shows the statistical significance of the difference between the ranking of each method and the 2$^{nd}$ best ranked method.  The ``Non. sig. (fail)'' column shows the fraction of all evaluation tests in which the alignment quality score is not statistically significant, and, in brackets, the fraction of evaluation tests in which the given NA method failed to produce an alignment.  Equivalent results over all evaluation tests broken down into functional and topological alignment quality measures, as well as over all evaluation tests broken down into network pairs/sets with known and unknown node mapping, are shown in Supplementary Tables \ref{tab:exp2_ranktable_ts_all_topo}--\ref{tab:exp1_ranktable_ts_unknown_all}. {\bf View II.} Alternative view of ranking of the NA methods.  Each pie chart shows the fraction of evaluation test ranks that fall into the 1--4, 5--8, and 9--12 rank bins out of all evaluation test ranks in the given alignment category. For example, for the PE framework, in the PE-P-P alignment category, 56\%, 26\%, and 18\% of the evaluation test ranks fall into ranks 1--4, 5--8, and 9--12, respectively, totaling to 100\% of the evaluation test ranks in the PE-P-P alignment category.  The pie charts allow us to compare the three alignment categories rather than individual NA methods in each category. The larger the pie chart for the better (lower) ranks, and the smaller the pie chart for the worse (higher) ranks, the better the alignment category. For example, in the PE framework, PE-P-P has the most evaluation tests ranked 1--4 and the fewest evaluation tests ranked 9--12, followed by PE-M-P, followed by PE-M-M. This implies that PE-P-P is superior to PE-M-P and PE-M-M.  The pie charts are color coded with respect to alignments of network pairs/sets with known and unknown node mapping, and TQ and FQ measures.  {\bf View III.} Overall ranking of an NA method versus its running time. The latter are running time results when aligning all network pairs in the Y2H$_1$ network set under the PE framework, and when aligning the Y2H$_1$ network set under the ME framework, where each method is restricted to use a maximum of 64 cores. The size of each point visualizes the overall ranking of the corresponding method over all evaluation tests over all network pairs/sets, corresponding to the ``Overall rank'' column in View I; the larger the point size, the better the method. In order to allow for easier comparison between the different alignment categories, ``Average'' shows the average running times and average rankings of the methods in each alignment category. Equivalent results where each method is restricted to use a single core are shown in Supplementary Figs. \ref{fig:exp5_pef} and \ref{fig:exp5_mef}. Equivalent results for T alignments are showing in Supplementary Fig. \ref{fig:res_t}. Detailed alignment quality scores are shown in Supplementary Tables \ref{tab:scores2} and \ref{tab:scores1}.
\pagebreak


\subsection{Method comparison focusing on accuracy of protein function prediction}
\label{sec:predsmain}

\subsubsection{New function prediction approach under the ME framework}
\label{sec:newpreds}

Here, we focus on addressing a potential issue with the existing approach for protein function prediction for multiple alignments, which we have used up to this point. As discussed in Section \ref{sec:prediction}, since the existing approach for multiple alignments (approach 2) is very different than the existing approach for pairwise alignments (approach 1), to make comparison between pairwise and multiple alignments (i.e., between PNA and MNA) more fair, we extend approach 1 for pairwise alignments into a new approach for multiple alignments (approach 3).

Then, we compare the new approach 3 against the existing approach 2, in hope that approach 3 will outperform  approach 2. If so, in our subsequent analyses, we will use approach 3 for protein function prediction for multiple alignments. This way, comparing results of approaches 1 and 3 will be much more fair than comparing results of approaches 1 and 2. Consequently, we will be able to more fairly compare PNA against MNA.

\begin{figure}[ht!]
\centering
\includegraphics[width=0.72\linewidth]{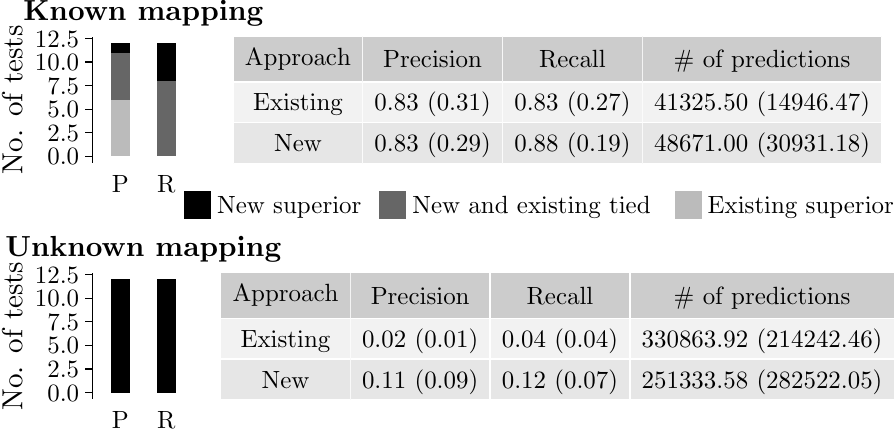}
\caption{Comparison of protein function prediction accuracy between the new (approach 3) versus existing (approach 2) prediction approach for multiple alignments.  We calculate prediction accuracy as follows. We apply the given approach (approach 2 or approach 3) to each alignment of each of the network sets from the ME framework, to predict protein-GO term associations. Then, we compute precision and recall for the given alignment's predicted protein-GO term associations.  Each bar on the left of the figure shows the number of tests (i.e., alignments) in which the new approach is superior, the existing approach is superior, or the two approaches are tied. Each table shows the precision, recall, and number of predictions averaged over all tests. \textcolor{black}{In parentheses, we show  standard deviations.} The results are separated into network sets with known and unknown node mapping.}
\label{fig:exp3_oldvsnew}
\end{figure}

Indeed, we find that our new approach 3 overall outperforms the existing approach 2 (Fig. \ref{fig:exp3_oldvsnew} and Supplementary Fig. \ref{fig:exp3_oldvsnew_full}). Specifically, approach 3 is overall comparable to approach 2 for networks with known node mapping (\textcolor{black}{marginally inferior} in terms of precision, marginally superior in  terms of recall) and it is superior to approach 2 for networks with unknown node mapping (in terms of both precision and recall). 

For networks with known node mapping, the number of predictions made by approach 3 is \textcolor{black}{just} 0.5\%-5.8\% \textcolor{black}{larger than} that made by approach 2, depending on the NA method, as shown in Supplementary Fig. S7 (with the exception of \textcolor{black}{ConvexAlign}, which produces up to 54\% more predictions under approach 3 than under approach 2). \textcolor{black}{The slightly more predictions by approach 3 could explain its slightly lower precision and slightly higher recall. But the differences in the number of predictions as well as accuracy of these two approaches  on networks with known mapping are so minor \textcolor{black}{(within 2\%-5\%)} that we consider them as comparable. }

For networks with unknown node mapping, the number of predictions made by approach 3 is 2\%-72\% smaller than the number of predictions made by approach 2, depending on the NA method (with exception of  \textcolor{black}{ConvexAlign and} BEAMS, which in one instance produce  \textcolor{black}{6\% and} 158\% more predictions, respectively, under approach 3). While the fewer predictions under approach 3 could explain higher precision of approach 3 compared to approach 2, interestingly, approach 3 also results in higher recall than approach 2, despite the latter making more predictions (Fig. 6). 
 
\subsubsection{Protein function prediction under PE versus ME
  frameworks}
\label{sec:predscompare}

Next, we compare protein function prediction accuracy between the PE and ME frameworks, relying on approach 1 for pairwise alignments and on the fairly comparable approach 3 for multiple alignments.

\begin{figure}[ht!]
\centering
\includegraphics[width=0.72\linewidth]{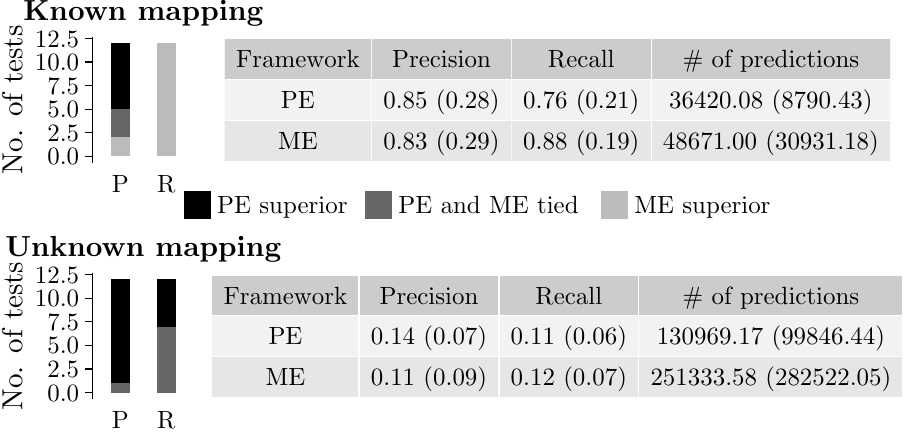}
\caption{Comparison of protein function prediction accuracy under the the PE and ME frameworks.  The figure can be interpreted the same way as Fig. \ref{fig:exp3_oldvsnew}. \textcolor{black}{Here,  we use new approach 3 for the ME framework. For analogous results when we use existing approach 2 for the ME framework, see Supplementary Fig. \ref{fig:exp3_pefvsmef_approach2}.} }
\label{fig:exp3_pefvsmef}
\end{figure}

For both the network sets with known and unknown node mapping, the predictions under the PE framework have higher precision while the predictions under the ME framework have higher recall (Fig. \ref{fig:exp3_pefvsmef} and Supplementary Fig. \ref{fig:exp3_pefvsmef_full}). Note that here, higher precision and lower recall for the PE framework compared to the ME framework could be due to somewhat fewer predictions under the PE framework than under the ME framework. Also, note that for networks with known node mapping, both sets of predictions have impressively high precision and recall scores, so any difference in their scores \textcolor{black}{(1\%-6\%)} can be considered marginal. This is not the case for networks with unknown node mapping, where the scores are lower. In this case, the superiority of the PE framework's precision over the ME framework's precision  \textcolor{black}{(17\%)} is more pronounced than the superiority of the ME framework's recall over the PE framework's recall \textcolor{black}{(8\%)}.  Additionally, achieving higher precision might be more preferred than achieving higher recall in the task of protein function prediction by experimental scientists who would potentially validate the predictions. Thus, we can argue that overall the PE framework (i.e., pairwise alignments) results in more accurate  predictions than the ME framework (i.e., multiple alignments). 



\section{Conclusion}
\label{sec:conclusion}

We introduce an evaluation framework for a fair comparison of PNA against MNA.  We find that (i) \textcolor{black}{the considered} PNA methods produce pairwise alignments that are superior to the corresponding pairwise alignments produced by \textcolor{black}{the considered} MNA methods, and (ii) \textcolor{black}{the} PNA methods produce multiple alignments that are superior to the corresponding multiple alignments produced by \textcolor{black}{the} MNA methods.  Also, using \textcolor{black}{the} pairwise alignments leads to higher protein function prediction accuracy than using \textcolor{black}{the} multiple alignments. Importantly, in addition to  PNA being overall more accurate, it is also overall faster than MNA.
This holds both both of T+S alignments and T alignments.

\textcolor{black}{In our evaluation, we have focused on comparing the two categories of approaches, PNA and MNA, rather than on identifying the best (PNA or MNA) approach. Some approaches may be better than others. 
In the PNA category, most of the considered approaches, and especially MAGNA++, perform  well consistently across the different scenarios (in both PE and ME framework, for both networks with known and unknown node mapping, and for both TQ and FQ), with some exceptions (Supplementary Tables \ref{tab:exp2_ranktable_ts_all_topo}--\ref{tab:exp1_ranktable_ts_unknown_all}). In the MNA category, only multiMAGNA++ works well consistently across all scenarios. Additionally, ConvexAlign works well for FQ and networks with unknown node mapping.}

\textcolor{black}{However, no method is always the best (i.e., has an overall rank of 1 over all evaluation tests). Namely, while in both PE and ME frameworks several PNA methods and the multiMAGNA++ MNA method achieve very good (low) overall ranks in the 1-2 range for networks with known node mapping or TQ, for networks with unknown node mapping and FQ, overall ranks start at about 4 (Supplementary Tables \ref{tab:exp2_ranktable_ts_all_topo}--\ref{tab:exp1_ranktable_ts_unknown_all}). That is, for networks with unknown mapping and FQ, even the best methods (ConvexAlign and multiMAGNA++) work well for some but not all networks or alignment quality measures. So, there seems to be a lot more room for improvement on how to better perform PNA or MNA to improve FQ (the quality of functional predictions) from networks with unknown mapping (PPI networks of different species). Fig. \ref{fig:exp3_pefvsmef} further signals this, given low prediction accuracy under both the PE and ME frameworks.}

\textcolor{black}{Importantly, the best approaches in our study in terms of FQ are of the one-to-one type, \textcolor{black}{which we hypothesize is because of heavier recent focus on and thus methodological advancements of such methods compared to those of the many-to-many type, per our discussion in Section \ref{sec:exp1}}. But one-to-one alignments cannot capture gene duplication events that exist in biological networks  \citep{Arabidopsis2011},  which  require existence of paralogs, i.e., a gene in one network being mapped to multiple genes in the same or another network. While many-to-many alignments can in theory capture these events, the considered many-to-many methods do not perform well in terms of FQ. So, developing better many-to-many methods might be a crucial future step in NA research. }

Since we demonstrate in the ME framework that  PNA can (by integrating  pairwise alignments) produce multiple alignments that are superior to multiple alignments produced by MNA, we believe that any new MNA methods should be compared not just to existing MNA methods but also   to existing PNA methods using our evaluation framework, to properly judge the quality of alignments that they produce. Our suggestion is similar to that of \citet{LocalVsGlobal}, who evaluated local versus global NA (rather than PNA versus MNA) and concluded that any new NA method should be compared against existing local as well as global NA methods.

Moreover, in the ME framework, PNA can produce multiple alignments that are superior to multiple alignments produced by MNA even with the simple variation of the pairwise alignment integration strategy (i.e., scaffolding procedure) introduced by \citet{SMAL}. Any more sophisticated scaffolding procedure that might be developed in the future will yield even more superior PNA-based multiple alignments and consequently even further emphasize the superiority of PNA over MNA. In other words, for MNA to gain advantage over PNA, a drastic redesign of the current MNA algorithmic principles \textcolor{black}{might} be needed.

\textcolor{black}{In summary, our current results suggest that perhaps it might be sufficient to focus on the faster PNA and integration of pairwise alignments into multiple ones rather than on the slower MNA. Of course, with development of newer approaches, the conclusions from our study might change. It is crucial that we (the NA community) gain in-depth understanding of practical implications of one-to-one versus many-to-many, pairwise versus multiple, local versus global, and other types of NA. This understanding is even more crucial given recent shift from traditional NA of static and homogeneous (single node type and single edge type) networks towards dynamic \citep{DynaMAGNA++,DynaWAVE,aparicio2019temporal} or heterogeneous \citep{HetNA1,HetNA2} NA.}


\paragraph{Funding.}
This work was supported by the Air Force Office of Scientific Research (AFOSR) [YIP FA9550-16-1-0147] and the National Institutes for Health (NIH) [1R01GM120733].



 
\bibliographystyle{apalike}

\newpage

\newcommand{\beginsupplement}{%
        \setcounter{section}{0}
        \renewcommand{\thesection}{S\arabic{section}}
        \setcounter{table}{0}
        \renewcommand{\thetable}{S\arabic{table}}%
        \setcounter{figure}{0}
        \renewcommand{\thefigure}{S\arabic{figure}}%
        \renewcommand{\figurename}{Supplementary Figure}

     }

\beginsupplement

\noindent\section*{Supplementary material: Pairwise versus multiple global network alignment}

\section{Methods}

\subsection{NA methods that we evaluate}
\label{supplsec:aligners}

The PNA methods that we evaluate are
GHOST, MAGNA++, WAVE, and L-GRAAL. The MNA methods that we
evaluate are IsoRankN, BEAMS, multiMAGNA++, \textcolor{black}{and ConvexAlign}.

\noindent{\bf PNA methods.}
Most NA methods are {\em two-stage} aligners: in the first stage, they
calculate the similarities (based on network topology and, optionally,
protein sequence information) between nodes in the compared
networks, and in the second stage, they use an alignment strategy to
find high scoring alignments with respect to the total similarity over
all aligned nodes.  GHOST is an example of two-stage PNA methods.
GHOST calculates the similarity of ``spectral signatures'' of nodes
between the compared networks in its first stage. Then, GHOST uses an
alignment strategy consisting of a seed-and-extend global alignment
step followed by a local search procedure that aims to improve, with
respect to node similarity, upon the seed-and-extend step.  An issue
with two-stage methods is that while they find high scoring alignments
with respect to total node similarity (a.k.a. node conservation), they
do not take into account the amount of conserved edges during the
alignment construction process. But the quality of a network alignment
is often measured in terms of the amount of conserved edges.  To address
this issue, MAGNA++ directly optimizes both edge and node conservation
{\it while} the alignment is constructed; its node conservation
measure typically uses graphlet-based node similarities
\citep{Milenkovic2008}.  MAGNA is a {\em
search-based} (rather than a two-stage) PNA method.  Search-based
aligners can directly optimize edge conservation or any other
alignment quality measure.  WAVE was proposed as a
two-stage (rather than search-based) PNA method that optimizes both a
graphlet-based node conservation measure as well as
(weighted) edge conservation by using a seed-and-extend alignment
strategy based on the principle of voting. Similarly, L-GRAAL
optimizes a graphlet-based node conservation measure and
a (weighted) edge conservation measure,
but it uses a seed-and-extend strategy based on integer programming.


\noindent{\bf MNA methods.}
IsoRankN is a two-stage MNA method.  It calculates node similarities between
all pairs of compared networks using a PageRank-based spectral
method. IsoRankN then creates a graph of the node similarities and
partitions the graph using spectral clustering in order to produce a many-to-many alignment.
BEAMS is a two-stage method that optimizes both a (protein sequence-based) node
conservation measure and
an edge conservation measure.  BEAMS uses a maximally weighted clique
finding algorithm on a graph of node similarities to produce a
one-to-one alignment, where node similarity is based only on protein
sequence information, without considering any topological node
similarity information.  BEAMS then creates a many-to-many alignment
from the one-to-one alignment using an iterative greedy algorithm that
maximizes both node and edge conservation. \textcolor{black}{ConvexAlign is also a two-stage method. It optimizes an objective function that combines topological node similarity, optional sequence-based node similarity, and edge conservation. That is, it optimizes both node and edge conservation. ConvexAlign optimizes its objective function with an optimization strategy that is formulated as an integer program, which is relaxed into a convex optimization problem. This problem is then solved using the alternating direction method of multipliers (ADMM). This allows ConvexAlign to  align multiple networks simultaneously.} Like MAGNA++, multiMAGNA++
is a search-based method that directly optimizes both edge and node
conservation while the alignment is constructed. \textcolor{black}{Of the MNA methods, IsoRankN and BEAMS produce many-to-many alignments, while ConvexAlign and multiMAGNA++
produce one-to-one alignments.}

\noindent{\bf Aligning using network topology only versus using both
  topology and protein sequences.}
In our analysis, for each method, we study the effect on output
quality when (i) using only network topology while constructing
alignments (T alignments) versus (ii) using both network topology and
protein sequence information while constructing alignments (T+S
alignments).  For T alignments, we set method parameters to ignore any
sequence information. All methods except BEAMS can produce T
alignments and all methods can produce T+S alignments.  For T+S
alignments, we set method parameters to include sequence information.
Supplementary Table \ref{tab:params} shows the specific parameters
that we use.  Specifically, the methods combine topological
information with sequence information in order to optimize
$\theta S_T + (1-\theta)S_P$, where $S_T$ is the (node or edge) cost
function based on {\em topological information}, $S_P$ is the node cost
function based on {\em protein sequence information}, and $\theta$ weighs
between topological information and sequence information.
When $\theta = 1$, only network topology is used in the alignment
process, and when $\theta = 0$, only sequence information is
used.  We set $\theta = 0.5$ in our study due to the following
reasons \textcolor{black}{(except for ConvexAlign, see below)}.
First, \citet{LocalVsGlobal}, who used the same datasets that we use in our
study, showed that as long as some amount of
topological information and some amount of protein sequence
information are used in the alignment process (i.e., as long as
$\theta$ does not equal 0 or 1), the quality of the resulting
alignments is not drastically affected.
They showed this for ten PNA methods, including GHOST, MAGNA++, WAVE,
and L-GRAAL, which are the PNA methods that we use in this study. Second,
it was shown by the original studies which introduced
two of the MNA methods used in this study that varying $\theta$ between 0.3 and
0.7 has no large effect on the quality of alignments produced by BEAMS
and IsoRank
\citep{BEAMS}, and that varying $\theta$ between 0.2 and 0.8 has no
large effect on the quality of alignments produced by FUSE
\citep{FUSE}. Third, the original
MAGNA++ paper, which multiMAGNA++ is based on, showed that varying
$\theta$ between 0.1 and 0.9 has no large effect on the quality of
alignments produced by MAGNA++. So, in the original multiMAGNA++ paper,
the $\theta$ parameter was set to 0.5. We believe
that all of this justifies our choice of using $\theta$ of 0.5 for
all methods considered in our study \textcolor{black}{(except for ConvexAlign, see below)}. Also, using the same $\theta$
value for all methods \textcolor{black}{(except for ConvexAlign, see below)} ensures that any potential differences in
results of the different methods are not caused by using different
amounts of network topology versus protein sequence information. \textcolor{black}{While in an ideal scenario we would have wanted to use $\theta=0.5$ for ConvexAlign's T+S alignments as well (just like we do for all other considered methods), the authors of ConvexAlign pre-set this value in ConvexAlign's implementation to a recommended value of 0.343 (see below), thus weighing topological information by 0.343 and sequence information by 0.657. We respect this recommendation and consequently use $\theta=0.343$ for ConvexAlign.}

\textcolor{black}{\textbf{Next, we clarify how the given method's parameter values from Supplementary Table \ref{tab:params} match the desired $\theta$ value.} }

\textcolor{black}{Recall that the methods combine topological information with sequence information
in order to optimize $\theta S_T + (1-\theta)S_P$, where $S_T$ is the
(node or edge) cost function based on {\em topological information},
$S_P$ is the node cost function based on {\em protein sequence
information}, and $\theta$ weighs between topological information and
sequence information.
}

\textcolor{black}{
\textbf{For T alignments}, we set parameters such that only topological
information is used (i.e., such that $\theta = 1.0$).  Namely, setting
$\theta = 1.0$ is equivalent to setting the following parameter value(s) for
each of the methods, where $E_T$, $N_T$, and $N_S$ are the
topological edge conservation function, topological node cost function, and sequence-based node cost function, respectively. (That is, $E_T$ and/or $N_T$ form $S_T$ from the above $\theta$-related formula, and $N_S$ is $S_P$ from the above $\theta$-related formula.)
}

\begin{itemize}

\item\textcolor{black}{
For GHOST, which optimizes $\alpha N_T + (1 - \alpha) N_S$, setting
$\theta = 1.0$ corresponds to setting $\alpha = 1.0$, i.e., alpha=1.0
in the GHOST implementation.
}

\item \textcolor{black}{
For L-GRAAL, which optimizes $(1-\alpha) E_T + \alpha N_S$ (where $E_T$ is edge conservation weighted by topological node similarity), setting
$\theta = 1.0$ corresponds to setting $\alpha = 0.0$, i.e., a=0.0 in
the L-GRAAL implementation.}

\item \textcolor{black}{
For MAGNA++, which optimizes $\alpha E_T + (1-\alpha)(\beta N_T +
(1-\beta) N_S)$, setting $\theta = 1.0$ corresponds to setting $\alpha
= 0.5$ and $\beta = 1.0$, i.e., setting a=0.5 and inputting only
topological node similarity into the MAGNA++ implementation, respectively. Note that we use a=0.5 to give equal weight to edge conservation and node conservation.}

\item \textcolor{black}{For WAVE, which optimizes $\alpha N_T + (1-\alpha) N_S$, setting $\theta = 1.0$ corresponds to setting $\alpha = 1.0$, i.e., inputting only topological node similarity to the WAVE implementation. Note that WAVE also optimizes edge conservation, but it does so implicitly, as a part of its alignment strategy. That is, edge conservation is not an input parameter of WAVE or its implementation.}

\item \textcolor{black}{
For IsoRankN, which optimizes $\alpha N_T + (1-\alpha) N_S$, setting
$\theta = 1.0$ corresponds to setting $\alpha = 1.0$, i.e., alpha=1.0
in the IsoRankN implementation.
}

\item \textcolor{black}{
For ConvexAlign, which optimizes $\lambda_2 E_T + (1-\lambda_2)
(\lambda_1 N_T + (1-\lambda_1) N_S)$, setting $\theta = 1.0$
corresponds to setting $\lambda_1 = 1.0$, i.e., inputting no node
similarity into the ConvexAlign implementation. Note that we use $\lambda_2$ of 0.02, as recommended and pre-set by the authors of the ConvexAlign paper. ConvexAlign authors have recommended all of its parameter values after testing them using cross-validation. So, we did not need to set any parameter values ourselves.
}

\item \textcolor{black}{
For multiMAGNA++, which optimizes $\alpha E_T + (1-\alpha)(\beta N_T +
(1-\beta) N_S)$, setting $\theta = 1.0$ corresponds to setting $\alpha
= 0.5$ and $\beta = 1.0$, i.e., setting a=0.5 and inputting only
topological node similarity into the multiMAGNA++ implementation, respectively. Note that we use a=0.5 to give equal weight to edge conservation and node conservation.
}

\end{itemize}

\textcolor{black}{
\textbf{For T+S alignments}, we set parameters such that both topological and
sequence information is used (i.e., such that $\theta = 0.5$, unless
recommended otherwise by the authors of the given method).  Namely, setting $\theta = 0.5$
is equivalent to setting the following parameter value(s) for each of the
methods.
}

\begin{itemize}

\item \textcolor{black}{
For GHOST, which optimizes $\alpha N_T + (1 - \alpha) N_S$, setting
$\theta = 0.5$ corresponds to setting $\alpha = 0.5$, i.e., alpha=0.5
in the GHOST implementation.
}

\item \textcolor{black}{
For L-GRAAL, which optimizes $(1-\alpha) E_T + \alpha N_S$ (where $E_T$ is edge conservation weighted by topological node similarity), setting
$\theta = 0.5$ corresponds to setting $\alpha = 0.5$, i.e., a=0.5 in
the L-GRAAL implementation.
}

\item \textcolor{black}{
For MAGNA++, which optimizes $\alpha E_T + (1-\alpha)(\beta N_T +
(1-\beta) N_S)$, setting $\theta = 0.5$ corresponds to setting $\alpha
= 0.25$ and $\beta = 0.33$,
i.e., setting a=0.25 and inputting the combined node similarity
information into the MAGNA++ implementation. With these parameter values, topological and sequence-based cost functions are equally
weighted. Namely, the optimization formula for MAGNA++ becomes $0.25 E_T + 0.75 (0.33 N_T + 0.67 N_S) = 0.25 E_T + 0.25 N_T + 0.5 N_S = 0.5 S_T + 0.5 S_P$, i.e., $\theta = 0.5$, as desired.
}

\item \textcolor{black}{For WAVE, which optimizes $\alpha N_T + (1-\alpha) N_S$, setting $\theta = 0.5$ corresponds to setting $\alpha = 0.5$, i.e., inputting both topological and sequence-based node similarities to the WAVE implementation. Note that WAVE also optimizes edge conservation, but it does so implicitly, as a part of its alignment strategy. That is, edge conservation is not an input parameter of WAVE or its implementation.}

\item \textcolor{black}{
For IsoRankN, which optimizes $\alpha N_T + (1-\alpha) N_S$, setting
$\theta = 0.5$ corresponds to setting $\alpha = 0.5$, i.e., alpha=0.5
in the IsoRankN implementation.
}

\item \textcolor{black}{
For ConvexAlign, which optimizes $\lambda_2 E_T + (1-\lambda_2)
(\lambda_1 N_T + (1-\lambda_1) N_S)$, we use $\lambda_1 = 0.33$ and $\lambda_2=0.02$, as recommended and pre-set by the authors of the ConvexAlign paper. ConvexAlign authors have recommended all of its parameter values after
testing them using cross-validation. So, we did not need to set any parameter values ourselves. With these two parameter values, the optimization formula for ConvexAlign becomes $0.02 E_T + 0.98 (0.33 N_T + 0.67 N_S) = 0.02 E_T + 0.323 N_T + 0.657 N_S = 0.343 S_T + 0.657 S_P$, i.e., $\theta = 0.343$. Clearly, ConvexAlign weighs sequence information higher than the other methods (65.7\% of the whole objective function for ConvexAlign, as opposed to 50\% of the while objective function for the other methods). Again, this is because the authors of ConvexAlign suggested doing 65.7\% for their method, while our justification for 50\% for the other methods is discussed above.
}

\item \textcolor{black}{
For multiMAGNA++, which optimizes $\alpha E_T + (1-\alpha)(\beta N_T +
(1-\beta) N_S)$, setting $\theta = 0.5$ corresponds to setting $\alpha
= 0.25$ and $\beta = 0.33$ as recommended by the multiMAGNA++ paper,
i.e., setting a=0.25 and inputting the combined node similarity
information into the multiMAGNA++ implementation. With these parameter values, topological and sequence-based cost functions are equally
weighted. Namely, the optimization formula for multiMAGNA++ becomes $0.25 E_T + 0.75 (0.33 N_T + 0.67 N_S) = 0.25 E_T + 0.25 N_T + 0.5 N_S = 0.5 S_T + 0.5 S_P$, i.e., $\theta = 0.5$. 
}

\end{itemize}

\subsection{Alignment quality measures}
\label{supplsec:quality}

Here, we describe the alignment quality measures that we use to
evaluate the NA methods. To do so, we first need to formally define an
alignment.
Typical PNA methods produce alignments comprising node pairs and
typical MNA methods produce alignments comprising node clusters.  We
introduce the term {\em aligned node group} to describe either an
aligned node pair or an aligned node cluster.  With this, we can
represent a pairwise or multiple alignment as a set of aligned node
groups.  Let $G_1(V_1,E_1)$, $\ldots$, $G_k(V_k,E_k)$ be $k$ networks
with node and edge sets $V_l$ and $E_l$, respectively, for
$l=1,2,\ldots,k$. An {\em alignment} of the $k$ networks is a set of
disjoint node groups, where each group is represented as a tuple
$(a_1,\ldots,a_k)$ with the following properties: (i) $a_l$ is the set
of nodes in the node group from network $G_l$, i.e., $a_l \subseteq
V_l$, for $l=1,2,\ldots,k$, (ii) no two node groups have any common
nodes, i.e., given two different groups $(a_1,a_2,\ldots,a_k)$ and
$(b_1,b_2,\ldots,b_k)$, $a_l \cap b_l = \varnothing$ for
$l=1,2,\ldots,k$, and (iii) there must be at least two nodes in each
node group, i.e., $\abs{\cup_{l=1,\ldots,k}{a_l}} \ge 2$.  If for each
node group in the given alignment there is at most one node from each
network, i.e., if for
each node group $\abs{a_l} \le 1$ for $l=1,\ldots,k$, then the
alignment is a {\em one-to-one} alignment. Otherwise, it is a {\em
many-to-many} alignment.

\subsubsection{Topological quality (TQ) measures}
\label{supplsec:topoquality}

A good NA method should produce aligned node groups that have internal
consistency with respect to protein labels. If we know the true node
mapping between the networks, then we can let the labels be protein
names. When the labels are based on the true node mapping, i.e., on
protein names, we
consider measures that rely on node labels to be capturing
topological alignment quality (TQ).  If we do not know the true node
mapping, we let the labels be GO terms. In this case, since GO terms
capture protein
functions, we consider measures that rely on GO terms to be
capturing functional alignment quality (FQ). We discuss such
measures in Supplementary Section \ref{supplsec:funcquality}.

Also, a good NA method should find a large amount of common network
structure across the compared networks, i.e., produce high edge
conservation.

Finally, for a good NA method, conserved edges should form large,
dense, connected regions (as opposed to small or isolated conserved
regions).

Below, first, we discuss how we measure internal consistency of aligned
protein groups in a pairwise alignment. Second, we comment on
how we do this in a multiple alignment. Third, we discuss how we
measure edge conservation in a pairwise alignment. Fourth, we comment
on how we do this in a multiple
alignment. Fifth, we discuss how we capture the notion of large,
dense, and connected conserved network regions (for both pairwise and
multiple alignments).

\vspace{0.25em}\noindent\textbf{1. Measuring internal node group
consistency of a pairwise alignment via precision, recall, and F-score
of node correctness (P-NC, R-NC, and F-NC, respectively)}. These measures \citep{LocalVsGlobal}
are a generalization of node correctness (NC) from one-to-one to
many-to-many pairwise alignments.  NC for one-to-one pairwise
alignments is the fraction of node pairs
from the alignment that are present in the true node mapping.
As such, NC evaluates the {\em
precision} of the alignment.  NC is extended to many-to-many
pairwise alignments as follows.  For each aligned node group $C_i$ in the
alignment, $C_i$ is converted into a set of
all possible $\abs{C_i} \choose 2$ node pairs in the group. The union of
all resulting node pairs over all groups $C_i$ forms the set $X$ of all aligned
node pairs.
Then, given
the set $Y$ of all node pairs from the true node mapping,
P-NC = $\frac{\abs{X \cap Y}}{\abs{X}}$, R-NC = $\frac{\abs{X \cap Y}}{\abs{Y}}$, and
F-NC is the harmonic mean of P-NC and R-NC.
These three
measures work for
both one-to-one and many-to-many pairwise alignments.

\vspace{0.25em}\noindent\textbf{2. Measuring internal node group consistency
of a multiple alignment via adjusted multiple node correctness
(NCV-MNC)}. Multiple node
correctness (MNC) \citep{multiMAGNA++} is a generalization of the NC
measure to multiple alignments. MNC uses the notion of
normalized entropy (NE), which measures, for a given aligned node group,
how likely it is to observe the same or higher level of internal node
group consistency
by chance, i.e., if the group of the same size was formed by randomly
assigning to it proteins from the compared networks.
The lower the NE, the more consistent the
node group.  Then, MNC is one minus
the mean of NEs across all node groups. We refer to
\citet{multiMAGNA++} for the formal definition of MNC.  Since a good NA
method should align (or cover) many of the nodes in the compared
networks, as was done by \citet{multiMAGNA++}, we adjust the MNC
measure to account for node coverage (NCV), which is the fraction of
nodes that are in the alignment out of all nodes in the compared networks.
Then, MNC-NCV$ = \sqrt{(\textrm{NCV})(\textrm{MNC})}$. When either NCV
or MNC is low, the geometric mean of the two is penalized.
The NCV-MNC measure
works for both one-to-one and many-to-many multiple
alignments.

\vspace{0.25em}\noindent\textbf{3. Measuring edge conservation of a
pairwise alignment via adjusted generalized S$^3$ (NCV-GS$^3$)}. Given
two compared networks, generalized S$^3$ (GS$^3$)
\citep{LocalVsGlobal} measures the fraction of conserved edges out of
both conserved and non-conserved edges, where an edge is conserved if
it maps to an edge in the other network and an edge is not conserved
if it maps to a non-adjacent node pair (i.e., a non-edge) in the other
network.  We refer to \citet{LocalVsGlobal} for formal definition of
GS$^3$.  As was done by \citet{LocalVsGlobal}, we penalize alignments
with low node coverage by combining NCV with GS$^3$ into the adjusted
GS$^3$ measure, NCV-GS$^3$, which equals
$\sqrt{(\textrm{NCV})(\textrm{GS}^3)}$.  The NCV-GS$^3$ measure works for both
one-to-one and many-to-many pairwise alignments.

\vspace{0.25em}\noindent\textbf{4. Measuring edge conservation of a multiple alignment via
adjusted cluster interaction quality (NCV-CIQ)}.
CIQ \citep{BEAMS} is a weighted sum of edge
conservation between all pairs of aligned node groups.  We refer to
\citet{multiMAGNA++} for the formal definition of CIQ. As was done by
\citet{multiMAGNA++}, we penalize alignments with low node coverage by
combining NCV with CIQ into the adjusted CIQ, NCV-CIQ, which equals
$\sqrt{(\textrm{NCV})(\textrm{CIQ})}$.
The NCV-CIQ measure
works for both one-to-one and
many-to-many multiple alignments.

\vspace{0.25em}\noindent{\bf 5. Measuring the size of the largest connected region
using largest common connected subgraph (LCCS).}  The LCCS measure, which was
recently extended from PNA \citep{MAGNA} to MNA \citep{multiMAGNA++},
simultaneously captures
the size (i.e., the number of nodes) and the density (i.e., the number of
edges) of the largest common connected subgraph
formed by the conserved edges, penalizing smaller or sparser
subgraphs. We refer to \citet{multiMAGNA++} for the formal definition
of LCCS.
The LCCS measure works
for both one-to-one and many-to-many alignments, and for both pairwise and multiple alignments.

\subsubsection{Functional quality (FQ) measures}
\label{supplsec:funcquality}

Per Supplementary
Section \ref{supplsec:topoquality}, a good alignment should have internally
consistent aligned node groups.  Instead of protein names as in Supplementary Section
\ref{supplsec:topoquality}, in this section we use GO terms as protein labels to
measure internal consistency.

Having aligned node groups that are internally consistent with respect to
protein labels is important for protein function prediction.
In addition to measuring internal node group consistency, we directly
measure the accuracy of protein function prediction.
That is, we first use a protein function prediction approach (Section
\ref{sec:prediction} of the main paper) to predict
protein-GO term associations, and then we compare the predicted
associations to known protein-GO term associations to see how accurate
the predicted associations are.

Below, first, we discuss how we measure internal node group
consistency with respect to GO terms. Second, we discuss an
alternative popular measure for
doing the same. Third, we discuss how we measure the
accuracy of protein function prediction, i.e., of predicted protein-GO
term associations (note that we describe a strategy that we use to
make the predictions in Section \ref{sec:prediction} of the main paper).

\vspace{0.25em}\noindent{\bf 1. Measuring internal node group
consistency using mean normalized entropy (MNE)}. MNE \citep{IsoRankN}
first uses normalized entropy (NE) to measure GO term-based
consistency of an individual aligned node group. The lower the NE, the
more consistent the given node group. Then, MNE is the mean of the NEs
across all node groups.  We refer to \citet{multiMAGNA++} for the
formal definition of MNE.  The MNE measure works for both one-to-one and
many-to-many alignments, and for both pairwise and multiple
alignments.

\vspace{0.25em}\noindent{\bf 2. Measuring internal node group
consistency using GO correctness (GC).}  GO correctness, which was
recently extended from PNA \citep{GRAAL} to MNA \citep{multiMAGNA++},
measures the internal consistency of aligned node groups with respect
to GO terms as follows.  For each node group $C_i$
in the alignment,
$C_i$ is converted into a set of all possible $\abs{C_i} \choose 2$ node
pairs in the group. The union of all resulting node pairs over all
groups $C_i$ forms the set $X$ of all aligned node pairs. A subset of $X$ that
consists of all node pairs in which each of the two nodes is annotated
with at least one GO term is denoted as $Y$.
Then, GO correctness is
the fraction of node pairs in $Y$ in which the two nodes are both
annotated with the same GO term. In other words, GO correctness is the
fraction of all pairs of aligned nodes in which the aligned nodes
share a GO term.
The GO correctness measure works for both
one-to-one and many-to-many alignments, and for both pairwise and
multiple alignments.

\vspace{0.25em}\noindent{\bf 3. Precision, recall, and F-score of protein
function prediction (P-PF, R-PF, and F-PF, respectively).}
We describe how we predict protein-GO term associations in Section
\ref{sec:prediction} of the main paper.
Here, we describe how we evaluate accuracy of such predictions. Given
predicted protein-GO term associations, we calculate accuracy of
the predictions via
precision, recall, and F-score measures. Formally, given the set $X$
of predicted protein-GO term associations, and the set $Y$ of known
protein-GO term associations, P-PF $ = \frac{\abs{X \cap Y}}{\abs{X}}$,
R-PF $ = \frac{\abs{X \cap Y}}{\abs{Y}}$, and F-PF is
the harmonic mean of precision and recall.
These three measures work for both
one-to-one and many-to-many alignments, and for both pairwise and
multiple alignments.

\subsubsection{Protein function prediction approaches}
\label{sec:suppl:ppf2}

\vspace{0.25em}\noindent{\bf Approach 3. New protein function
  prediction for
  multiple alignments.}
We follow our discussion from Section \ref{sec:prediction} of the main
paper regarding
approach 3, our new protein function
prediction approach for multiple alignments.
Formally, given an alignment of
$k$ networks, $G_1(V_1,E_1)$, $G_2(V_2,E_2)$, $\ldots$, $G_k(V_k,E_k)$, and
given node $v$ in the alignment that has at least one annotated GO
term, and given GO term $g$, we hide the
protein's true GO term(s) and find the significance of the alignment
with respect to GO term $g$ using the hypergeometric test, as follows.
For each node group $C_i$ in the alignment, we convert $C_i$
into a set of node pairs $F_i$ by taking all node pairs in the node
group, after which we concatenate the sets of node pairs into a single
set $F$.
Then, let $V_i^*
\subset V_i$ be such that each node in $V_i^*$ is annotated with at
least one GO term.  Let $S_1$ be the set of all possible pairs of
proteins in $F$ such that one protein is in $V_i^*$ and the other is
in $V_j^*$, where $i \ne j$.  Let $A_i \subset V_i^*$ be such that
each node in $A_i$ is annotated with $g$.  Let $S_2$ be the set of all
possible pairs of proteins between $A_i$ and $A_j$, where $i \ne j$.
Let $K$ be the set of pairs of proteins that are in $F$ and in $S_1$.
Let $X$ be the set of pairs of proteins that are in $F$ and in $S_2$.
Then, we use the hypergeometric test to calculate the probability of
observing by chance $\abs{X}$ or more pairs of proteins in $F$ with
each node annotated with $g$ is $p = 1 -
\sum_{i=0}^{\abs{X}-1}\frac{{\abs{K} \choose i}{\abs{S_1}-\abs{K}
\choose \abs{S_2}-i}}{{\abs{S_1} \choose \abs{S_2}}}.$ We consider the
alignment to be significant with respect to $g$ if the $p$-value is
less than 0.05.  We predict $v$ to be associated with $g$ if the
alignment is significant with respect to $g$, resulting in predicted
protein-GO term associations.  If the alignment is significant with
respect to $g$, we predict $v$ to be associated with $g$. Repeating
this process for all nodes and GO terms results in predicted
protein-GO term associations $X$.

\subsubsection{Statistical significance of alignment quality scores}
\label{supplsec:significance}

We continue our discussion from Section \ref{sec:significance} of the
main paper on how
to compute the $p$-value of a quality score of an actual
alignment. This is done as
follows. We construct a set of 1,000 corresponding random alignments
(1,000 is what was practically reasonable given our computational
resources), under a null model that conserves the following properties
of the actual alignment: the number of node groups, the number of
nodes in each group, and the network from which each node in each node
group originates from.  Then, the $p$-value of the alignment quality
score is the fraction of the 1,000 random alignments with equal or
better score than the actual alignment.  We consider an alignment
quality score to be significant if its $p$-value is less than
$0.001$. Note that if a given method fails to produce an alignment of
a network pair/set, we set the $p$-values of all quality scores
associated with the method and network pair/set to 1 and hence
consider all of the associated quality scores to be non-significant.

\subsection{Evaluation framework}
\label{supplsec:evaluation}

\subsubsection{Multiple evaluation (ME) framework}
\label{supplsec:mef}

We continue our discussion from Section \ref{sec:mef} of the main paper on how we combine
the pairwise alignments over every network pair in the given set into
a multiple alignment, i.e., how we produce alignments from the ME-P-P
and ME-M-P categories. This procedure was inspired by \citet{SMAL}.
Given pairwise alignments of $k$ networks
$G_1(V_1,E_1),\ldots,G_k(V_k,E_k)$, \citet{SMAL} produce a multiple
alignment of the $k$ networks as follows.  First, they select a
``scaffold'' network $G_r$ among the $k$ networks, namely the network
whose sum of ``topological similarities'' to the remaining $k-1$
networks is maximized; one of the suggested ``topological similarity''
measures is Graphlet Degree Distribution (GDD) agreement
\citep{GDD}. Second, they align $G_r$ to
each of the remaining $k-1$ networks. Third, they take the union of
all aligned node groups from the resulting $k-1$ alignments. Let us
denote this union as set $A$.
Since the node groups in
set $A$ are not necessarily disjoint, \citet{SMAL} use set $A$ to
create a new set $A'$ of aligned node groups that are disjoint. This
is done as follows. Let $A'$ be an empty set. First, randomly pick an
aligned node group $C$ that is currently in $A$ (initially, all node
groups are in $A$) and remove it from $A$. Then, remove from $A$ all
node groups that have at least one node in
common with $C$, and merge the node groups into $C$.
Repeat this process until there are no more node groups in $A$ that have at
least one node in common with $C$. Then, add $C$ to $A'$.
Repeat this process until $A$ is
empty. This results in a new multiple alignment $A'$.  We illustrate
this procedure in Fig. \ref{fig:pemm}(b,c) of the main paper.  In our
work, instead of choosing one of the $k$ analyzed networks
as a scaffold network using BLAST protein similarity information as
\citet{SMAL} does, because the choice of scaffold network affects the
quality of the resulting multiple alignment (which we actually
validate in Supplementary Fig. \ref{fig:scaf}), we vary each of the
$k$ networks as the scaffold network $G_r$, and we choose the scaffold
based on the quality of the resulting multiple alignments. 
That is, we rank (as explained below) each of the $k$ multiple
alignments, in order to select the best (in terms of the rank) of
them. We rank the alignments as follows. For each
alignment quality measure, we rank the alignments from the best one to
the worst one. (In case of ties, we let the ranks of the tied
alignments be the tied alignments' average rank.) Then, we compute the
total rank of each alignment by taking the average of the given
alignment's ranks over all of the alignment quality measures. Finally,
we select the best (in terms of the total rank) of all alignments.
Note that here, we consider all measures that can deal with multiple
alignments, except NCV-MNC, which we leave out because not all network
pairs/sets have the true node mapping (and NCV-MNC requires knowing
this mapping), and except MNE, which we leave
out so that the number of TQ and FQ measures matches (which is
required in order to prevent the ranks to be dominated by topological
or functional alignment quality). That is, we
consider NCV-CIQ and LCCS TQ measures and GC and F-PF FQ measures.

\subsection{T versus T+S alignments}\label{sec:suppl_tvsts}

\begin{figure}[h!]
\centering
\includegraphics[width=0.7\linewidth]{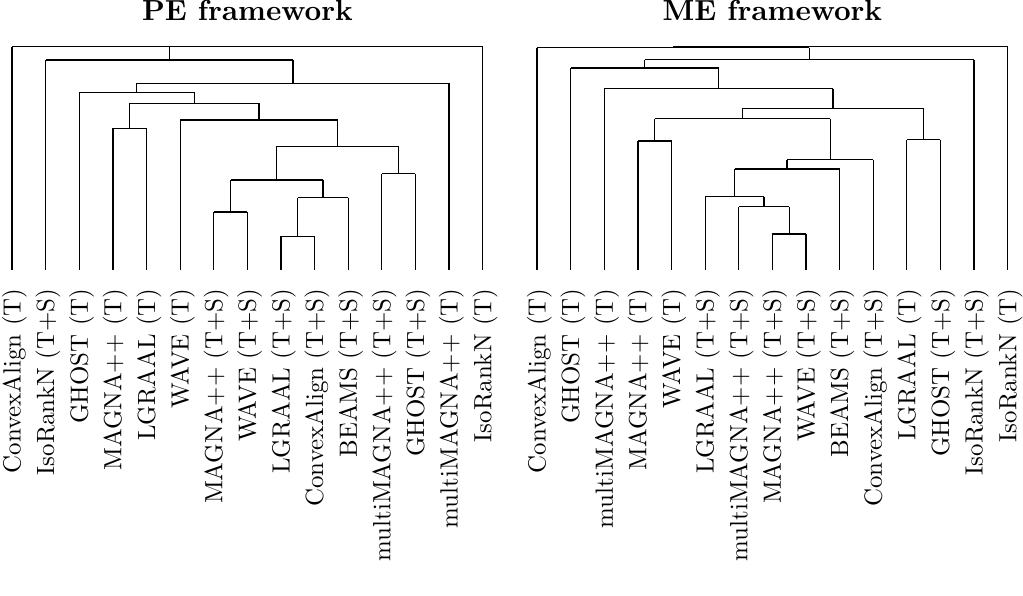}%
\caption{Clustering of NA methods, each with its T and T+S versions,
using each of the {\bf PE} and {\bf ME} frameworks.  Clustering is based on
pairwise method similarities, which we compute as follows.  The
similarity between two NA methods is the mean of the Adjusted Rand
Index (ARI; explained below) of each pair of corresponding alignments
produced by the two NA methods, over all network pairs/sets.  Each
alignment of a network pair/set is a set of node groups, i.e., a
partition of the nodes in all of the networks in the network pair/set,
and we measure similarity between two alignments by comparing their
partitions using ARI.  ARI \citep{PartitionSimilarity} is a widely
used measure to calculate the similarity between two partitions. Given
the similarities between all pairs of the NA methods, we cluster using
complete linkage hierachical clustering \citep{ClusterAnalysis} and
visualize the clustering using a dendrogram.  The results shown in
this figure rely on all alignments over all network sets (Yeast+\%LC,
PHY$_1$, PHY$_2$, Y2H$_1$, and Y2H$_2$).  Equivalent results broken
down into results for networks with known node mapping and results for
networks with unknown node mapping are shown in Supplementary Figs.
\ref{fig:exp2_dendro} and \ref{fig:exp1_dendro}, respectively.  }
\label{fig:expboth_dendro}
\end{figure}

\begin{figure}[h!]
\centering
{\bf(a)}\includegraphics[width=0.43\linewidth]{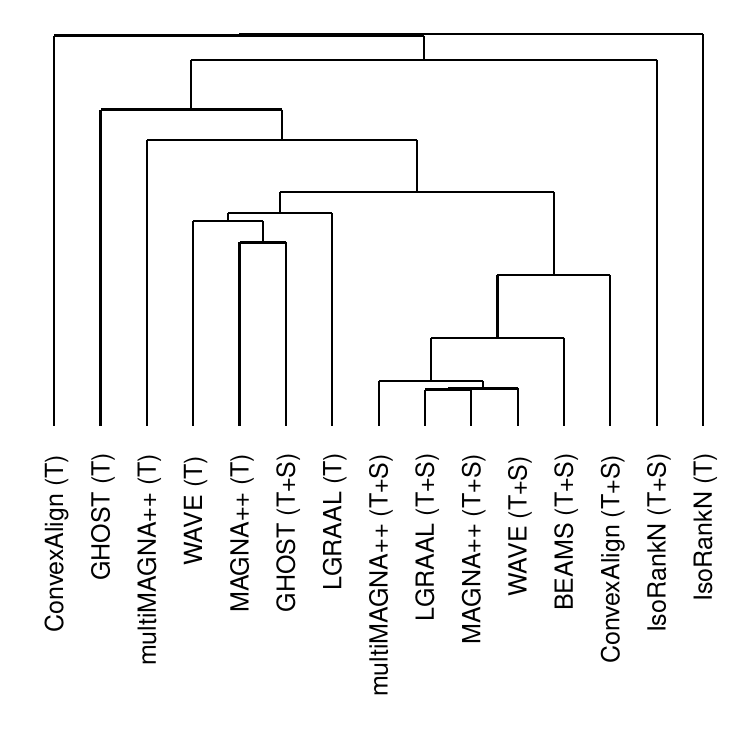}
{\bf(b)}\includegraphics[width=0.43\linewidth]{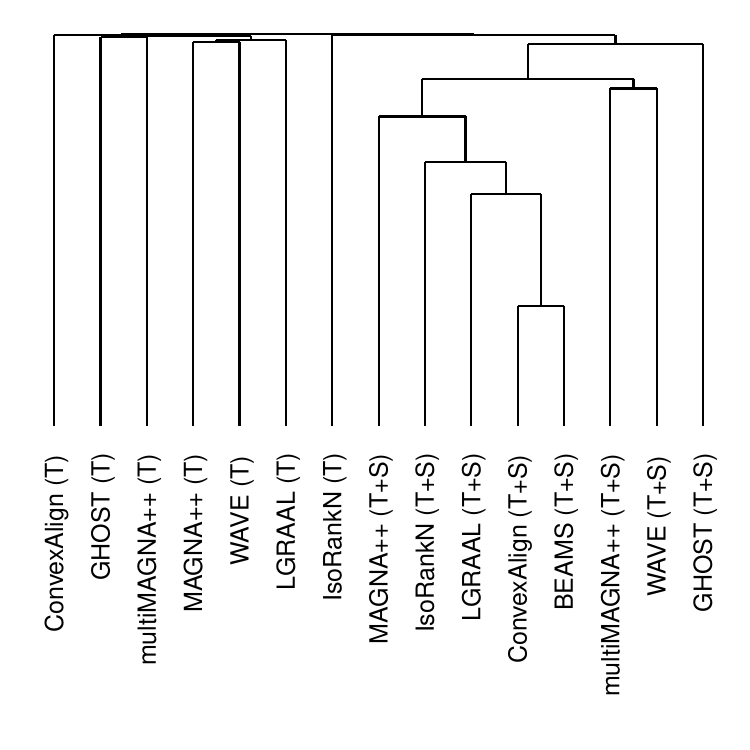}
\caption{ Clustering of NA methods, each with its T and T+S versions,
using all network sets with (a) {\bf known node mapping} and (b) {\bf
unknown node mapping} in the {\bf PE framework}.  The figure can be
interpreted the same way as Supplementary
Fig. \ref{fig:expboth_dendro}.  }
\label{fig:exp2_dendro}
\end{figure}

\begin{figure}[h!]
\centering
{\bf(a)}\includegraphics[width=0.43\linewidth]{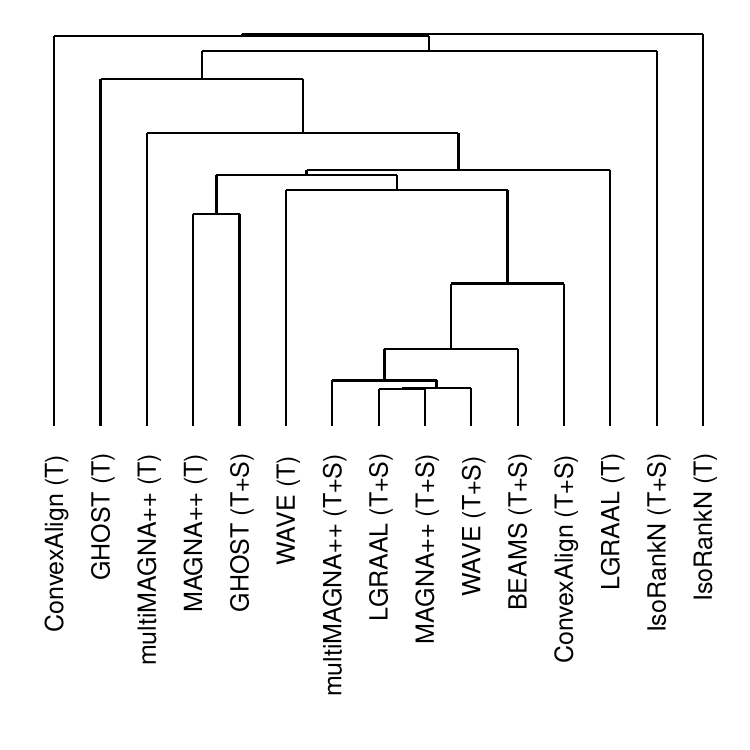}
{\bf(b)}\includegraphics[width=0.43\linewidth]{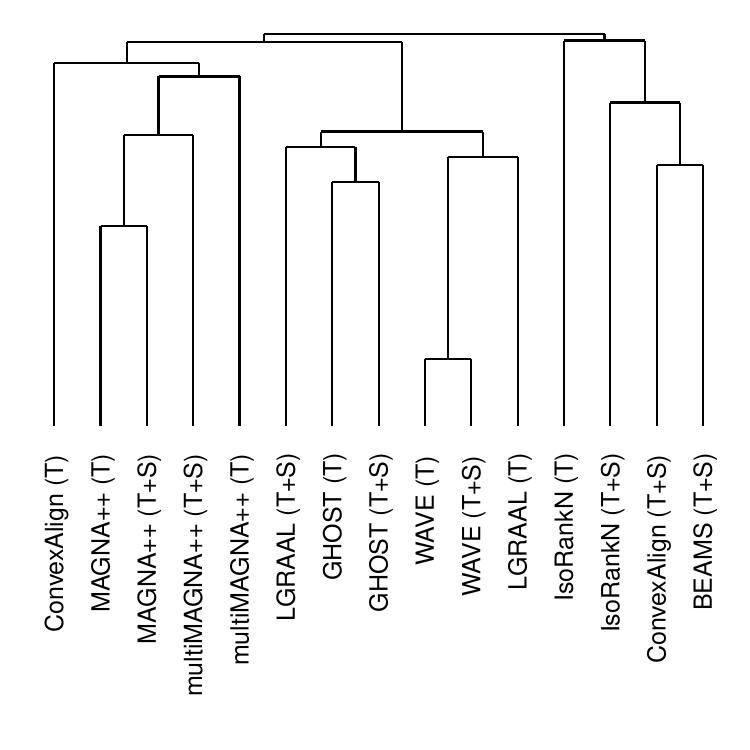}
\caption{ Clustering of NA methods, each with its T and T+S versions,
using all network sets with (a) {\bf known node mapping} and (b) {\bf
unknown node mapping} in the {\bf ME framework}.  The figure can be
interpreted the same way as Supplementary
Fig. \ref{fig:expboth_dendro}.  }
\label{fig:exp1_dendro}
\end{figure}

\textcolor{black}{Here, we continue our discussion from Section \ref{sec:tvsts} of the main paper regarding the similarity (overlap) of the alignments produced the
 different NA methods, each with its T and T+S versions (Supplementary
 Figs. \ref{fig:expboth_dendro}--\ref{fig:exp1_dendro}).  Surprisingly, \textcolor{black}{over all considered network datasets,
in each of the PE and ME frameworks,} the T+S versions of the different
 methods are overall more similar than the T+S and T versions of the same
 method are \textcolor{black}{(with the exception of IsoRankN in the PE framework and also GHOST in the ME framework)}.  That is, the T+S versions of the different methods
 cluster together in Supplementary Fig.
\ref{fig:expboth_dendro} and are clearly 
separated from the T versions. In contrast, the T versions do not cluster together.  This shows that using protein sequence information overall yields alignment consistency independent of which NA method is used. \textcolor{black}{Similar holds when we break down this analysis for networks with known versus unknown node mapping (Supplementary Figs. \ref{fig:exp2_dendro}--\ref{fig:exp1_dendro}), with the exception of networks with unknown node mapping under the ME framework, where the T and T+S versions of the same approach are often clustered together.  }}

\subsection{Method comparison in the ME framework: accuracy versus running time}\label{sect:suppl_ME_runtime}

\textcolor{black}{The running time discussion in Section \ref{sec:exp1} of the main paper deals with empirical running times of the
considered PNA and MNA methods, when the methods are run on our
considered network sets, the largest one of which contains six
networks. Since the PNA methods must align every pair of networks in a
network set in order to produce a multiple alignment, and since this
results in a quadratically increasing running time with respect to the
number of networks $k$, we next ask whether there is some (larger than
six) value of $k$ at which PNA might become less efficient (i.e.,
slower) than MNA. To answer this, because of the limited sizes (in
terms of $k$) of our considered network sets, we need to analyze the
methods' theoretic running time complexities with respect to $k$
 (Supplementary Table \ref{tab:complexity}).
All of the PNA methods's running times grow quadratically with $k$
due to the required pairwise alignments, per the above discussion.
Of the MNA methods, IsoRankN's running time also grows
quadratically, \textcolor{black}{ConvexAlign's running time grows cubically,} BEAMS' running time grows exponentially, and
multiMAGNA++'s time grows linearly with $k$. So, when comparing the
PNA and MNA methods, only multiMAGNA++ grows slower (i.e., is expected
to be faster) with $k$ than the PNA methods, IsoRankN  grows
at the same rate as the PNA methods, and \textcolor{black}{ConvexAlign and} BEAMS grow faster than the PNA methods. Hence, we do not expect that the MNA
methods will have advantage over the PNA methods as $k$ increases,
with the exception of multiMAGNA++. However, note that the analysis of
the methods' theoretic running times is different than the analysis of
their empirical running times, and also, note that their theoretic as
well as empirical running times depend not just on $k$ but also on the
sizes of the considered networks in terms of the numbers of their
nodes and edges, and also potentially on some method-specific
parameters. For example, while multiMAGNA++ theoretically grows the
slowest with $k$ of all considered PNA and MNA methods, as we can see
from its empirical running time analysis (Fig. \ref{fig:res}, View III, in the main paper),
multiMAGNA++ is significantly slower than BEAMS on
our considered network sets with up to six networks. So, it is hard to
estimate the exact value of $k$ at which multiMAGNA++ would
empirically perform faster than the other methods, as this would also
depend on e.g., the size of each network in the considered network
set. }

\clearpage
\newpage


\section*{SUPPLEMENTARY TABLES}

\begin{table}[h!]
\begin{center}
\begin{tabular}{llrr}
\toprule
Set & Species & No. of proteins & No. of interactions \\ \midrule
\multirow{6}{*}{Yeast+\%LC}
&                      Yeast+0\%LC   & 1,004 &  8,323 \\
&                      Yeast+5\%LC   & 1,004 &  8,739 \\
&                      Yeast+10\%LC  & 1,004 &  9,155 \\
&                      Yeast+15\%LC  & 1,004 &  9,571 \\
&                      Yeast+20\%LC  & 1,004 &  9,987 \\
&                      Yeast+25\%LC  & 1,004 & 10,403 \\ \midrule
\multirow{4}{*}{PHY$_1$}
&                       Fly     & 7,887 & 36,285     \\
&                       Worm    & 3,006 & 5,506      \\
&                       Yeast   & 6,168 & 82,368     \\
&                       Human   & 16,061 & 157,650   \\ \midrule
\multirow{2}{*}{PHY$_2$}
&                       Yeast   & 768 & 13,654     \\
&                       Human   & 8,283 & 19,697   \\ \midrule
\multirow{4}{*}{Y2H$_1$}
&                       Fly     & 7,097 & 23,370   \\
&                       Worm    & 2,874 & 5,199    \\
&                       Yeast   & 3,427 & 11,348   \\
&                       Human   & 9,996 & 39,984   \\ \midrule
\multirow{2}{*}{Y2H$_2$}
&                       Yeast   & 744 & 966       \\
&                       Human   & 1,191 & 1,567   \\ \bottomrule
\end{tabular}
\normalsize
\end{center}
\caption{Details on the PINs that we use in our study.
The true node mapping is known for the Yeast+\%LC network  set, unlike
for  the PHY$_1$, PHY$_2$, Y2H$_1$,
  and Y2H$_2$ network sets. Since the largest connected components of the
fly and worm networks in PHY$_2$ and Y2H$_2$ are too small, we do not use
those networks in our analysis.}
\label{tab:nets}
\end{table}

\newgeometry{left=0.5in,right=0.5in}
\begin{table}[h!]
\begin{center}
\scriptsize 
\begin{tabular}{ll}
\toprule
Algorithms & Parameters  \\ \midrule
\multicolumn{2}{l}{PNA methods, T alignments} \\ \midrule
GHOST      & beta=1e10 alpha=1.0 \\
L-GRAAL    & a=0.0 node similarity = graphlet degree vector (GDV) similarity\\
MAGNA++    & m=S3 p=15000 n=10000 a=0.5 node similarity = GDV similarity\\
WAVE & node similarity = GDV similarity \\
\midrule
\multicolumn{2}{l}{PNA methods, T+S alignments} \\ \midrule
GHOST      & beta=1e10 alpha=0.5 \\
L-GRAAL         & a=0.5 node similarity = GDV and BLAST protein similarity\\
MAGNA++       & m=S3 p=15000 n=10000 a=0.25 node similarity = GDV and BLAST protein similarity\\
WAVE &  node similarity = GDV and BLAST protein
       similarity\\ \midrule
\multicolumn{2}{l}{MNA methods, T alignments} \\ \midrule
IsoRankN       & K=30 thresh=1$\text{e-}$4 maxveclen=5000000 alpha=1.0\\
ConvexAlign  & lambda\_edge=3 numOuterIterations=4 flag\_fast=1 mu=150 min\_val=0.5 lambda\_mul=0.5 node similarity = none\\
multiMAGNA++    & m=CIQ p=15000 n=100000 e=0.5 a=0.5 node similarity = GDV similarity  \\ \midrule
\multicolumn{2}{l}{MNA methods, T+S alignments} \\ \midrule
IsoRankN      & K=30 thresh=1$\text{e-}$4 maxveclen=5000000 alpha=0.5
                node similarity = BLAST protein similarity   \\
ConvexAlign  & lambda\_edge=3 numOuterIterations=4 flag\_fast=1 mu=150 min\_val=0.5 lambda\_mul=0.5 node similarity = BLAST protein similarity\\                 
BEAMS         & beta=0.4 alpha=0.5 node similarity = BLAST protein similarity\\
multiMAGNA++  & m=CIQ p=15000 n=100000 e=0.5 a=0.25 node similarity = GDV and BLAST protein similarity\\ \bottomrule
\end{tabular}

\normalsize
\end{center}
\caption{
\textcolor{black}{Method parameters that we use in our study.  We use parameters
recommended in the methods' original publications.  The parameter
``node similarity'' indicates the node similarity information that is
inputted to the NA method. Note that graphlet degree vector (GDV)
similarity uses only network topological information, while BLAST
protein similarity uses only protein sequence information. Note that sometimes  different methods
use the same name (e.g., $\alpha$) for different parameters, or they use different
names (e.g., $\alpha$ versus $a$) for the same parameter.  For T
alignments, we set parameters such that only topological information
is used (i.e., such that $\theta = 1.0$; see Supplementary Section
\ref{supplsec:aligners}).  For T+S alignments, we set parameters
such that topological and sequence information are equally weighted
(i.e., such that $\theta = 0.5$; see Supplementary Section
\ref{supplsec:aligners}), as recommended by \citet{LocalVsGlobal}.
The only exception is ConvexAlign, for which we use a lower $\theta$ value, as recommended and pre-set in its implementation by its authors.  See Supplementary Section
\ref{supplsec:aligners} for details. }
}
\label{tab:params}
\end{table}
\restoregeometry

\begin{table}[h!]
\begin{center}
\begin{tabular}{ll} 
\toprule
Algorithms & Time complexity  \\ \midrule
\multicolumn{2}{l}{PNA methods} \\ \midrule
  GHOST         & $O(n (\frac{m}{n})^4) = O(\frac{m^4}{n^3})$ \\
  L-GRAAL & $O(n^3 + n^2 \frac{m}{n}^3) = O(n^3 + \frac{m^3}{n})$ \\
MAGNA++       & $O(n + m)$\\
WAVE          & $O(n^3)$\\ \midrule
\multicolumn{2}{l}{MNA methods} \\ \midrule
IsoRankN      & $O({k \choose 2} m^2) = O(k^2 m^2)$\\
ConvexAlign        & $O(k^3n^3)$\\
BEAMS         & $O(k^2 n^2 (\frac{m}{n})^{k+1})$\\
multiMAGNA++  & $O(k (n + m))$\\ \bottomrule
\end{tabular}
\normalsize
\end{center}
\caption{Theoretic time complexity of the considered PNA methods when
  they align two networks and of the considered MNA methods when they
  align $k$ networks, with respect to network size and the number of
  networks. Regarding network size, $n$ and $m$ is the number of nodes
  and edges, respectively, averaged over all networks under
  consideration.}
\label{tab:complexity}
\end{table}

\begin{table}[h!]
\centering
\begin{tabular}{rrrrr}
  \toprule
NA method & Overall rank & $p_1$-value & $p_2$-value & Non-sig (fail) \\ 
  \midrule
\cellcolor{green!30}{WAVE (PE-P-P)} & 1.70 (1.23) & NA & NA & 0.00 (0.00) \\ 
  \cellcolor{red!30}{multiMAGNA++ (PE-M-P)} & 1.93 (1.32) & 2.28e-01 & NA & 0.00 (0.00) \\ 
  \cellcolor{green!30}{MAGNA++ (PE-P-P)} & 3.21 (1.85) & 1.39e-04 & 2.64e-06 & 0.00 (0.00) \\ 
  \cellcolor{green!30}{GHOST (PE-P-P)} & 4.09 (3.66) & 1.06e-04 & 7.27e-04 & 0.14 (0.14) \\ 
  \cellcolor{green!30}{LGRAAL (PE-P-P)} & 4.21 (2.18) & 5.13e-08 & 2.00e-06 & 0.05 (0.05) \\ 
  \cellcolor{blue!30}{multiMAGNA++ (PE-M-M)} & 5.09 (1.56) & 2.36e-07 & 6.92e-08 & 0.00 (0.00) \\ 
  \cellcolor{red!30}{BEAMS (PE-M-P)} & 8.74 (1.99) & 5.06e-09 & 6.81e-09 & 0.02 (0.00) \\ 
  \cellcolor{blue!30}{ConvexAlign (PE-M-M)} & 9.07 (1.56) & 5.23e-09 & 5.19e-09 & 0.00 (0.00) \\ 
  \cellcolor{red!30}{ConvexAlign (PE-M-P)} & 9.09 (2.27) & 5.07e-09 & 5.02e-09 & 0.00 (0.00) \\ 
  \cellcolor{blue!30}{BEAMS (PE-M-M)} & 9.16 (1.90) & 5.08e-09 & 6.59e-09 & 0.09 (0.00) \\ 
  \cellcolor{red!30}{IsoRankN (PE-M-P)} & 9.56 (1.75) & 4.80e-09 & 4.45e-09 & 0.23 (0.00) \\ 
  \cellcolor{blue!30}{IsoRankN (PE-M-M)} & 9.63 (2.21) & 4.89e-09 & 4.65e-09 & 0.33 (0.00) \\ 
   \bottomrule

\end{tabular}
\vspace{0.15cm}
\caption{Overall ranking of the NA methods for the {\bf PE framework}
over all evaluation tests (where a test is a combination of an NA
method, a network pair, and an alignment quality measure) that use
{\bf TQ measures}, for T+S alignments, for networks with both known
and unknown node mapping.  By NA method, here, we
mean the combination of a PNA or MNA method and the alignment category
(Section \ref{sec:evaluation} of the main paper). Namely, there are 12 NA methods in the
PE framework (four PNA methods associated with the PE-P-P categories
and four MNA methods associated with each of the PE-M-M and PE-M-P
categories). The alignment categories are color coded.  The ``Overall
rank'' column shows the rank of each method averaged over all
evaluation tests, along with the corresponding standard deviation (in
brackets). Since there are 12 methods in a given framework, the
possible ranks range from 1 to 12. The lower the rank, the better the
given method.  The ``$p_1$-value'' column shows the statistical
significance of the difference between the ranking of each method and
the 1$^{st}$ best ranked method.  The ``$p_2$-value'' column shows the
statistical significance of the difference between the ranking of each
method and the 2$^{nd}$ best ranked method.  The
``Frac. non. sig. (failed)'' column shows the fraction of evaluation
tests in which the alignment quality score is not statistically
significant, and, in brackets, the fraction of evaluation tests in
which the given NA method failed to produce an alignment.}
\label{tab:exp2_ranktable_ts_all_topo}
\end{table}

\begin{table}[h!] 
\centering
\begin{tabular}{rrrrr}
  \toprule
NA method & Overall rank & $p_1$-value & $p_2$-value & Non-sig (fail) \\ 
  \midrule
\cellcolor{red!30}{ConvexAlign (PE-M-P)} & 4.33 (4.25) & NA & NA & 0.07 (0.00) \\ 
  \cellcolor{green!30}{MAGNA++ (PE-P-P)} & 4.98 (3.43) & 3.32e-01 & NA & 0.11 (0.00) \\ 
  \cellcolor{blue!30}{ConvexAlign (PE-M-M)} & 5.42 (4.60) & 4.29e-02 & 2.60e-01 & 0.21 (0.00) \\ 
  \cellcolor{red!30}{multiMAGNA++ (PE-M-P)} & 6.14 (4.17) & 7.71e-02 & 4.62e-05 & 0.19 (0.00) \\ 
  \cellcolor{green!30}{LGRAAL (PE-P-P)} & 7.02 (3.82) & 2.40e-03 & 1.10e-03 & 0.30 (0.05) \\ 
  \cellcolor{green!30}{WAVE (PE-P-P)} & 7.21 (4.21) & 7.42e-03 & 6.25e-07 & 0.25 (0.00) \\ 
  \cellcolor{blue!30}{IsoRankN (PE-M-M)} & 7.51 (3.48) & 2.34e-05 & 5.87e-04 & 0.28 (0.00) \\ 
  \cellcolor{blue!30}{multiMAGNA++ (PE-M-M)} & 7.54 (4.22) & 1.55e-03 & 3.80e-05 & 0.35 (0.00) \\ 
  \cellcolor{green!30}{GHOST (PE-P-P)} & 7.56 (4.33) & 2.56e-03 & 3.61e-06 & 0.37 (0.16) \\ 
  \cellcolor{red!30}{IsoRankN (PE-M-P)} & 7.82 (4.08) & 4.04e-05 & 6.55e-05 & 0.39 (0.00) \\ 
  \cellcolor{red!30}{BEAMS (PE-M-P)} & 8.33 (4.28) & 1.50e-05 & 1.77e-05 & 0.39 (0.00) \\ 
  \cellcolor{blue!30}{BEAMS (PE-M-M)} & 8.79 (4.22) & 8.96e-06 & 2.30e-06 & 0.47 (0.00) \\ 
   \bottomrule

\end{tabular}
\vspace{0.15cm}
\caption{Overall ranking of the NA methods for the {\bf PE framework}
over all evaluation tests (where a test is a combination of an NA
method, a network pair, and an alignment quality measure) that use
{\bf FQ measures}, for T+S alignments, for networks with both known
and unknown node mapping.  By NA method, here, we
mean the combination of a PNA or MNA method and the alignment category
(Section \ref{sec:evaluation} of the main paper). Namely, there are 12 NA methods in the
PE framework (four PNA methods associated with the PE-P-P categories
and four MNA methods associated with each of the PE-M-M and PE-M-P
categories).
The table can be interpreted the same way as Supplementary Table \ref{tab:exp2_ranktable_ts_all_topo}.}
\label{tab:exp2_ranktable_ts_all_func}
\end{table}

\begin{table}[h!]
\centering
\begin{tabular}{rrrrr}
  \toprule
NA method & Overall rank & $p_1$-value & $p_2$-value & Non-sig (fail) \\ 
  \midrule
\cellcolor{red!30}{multiMAGNA++ (PE-M-P)} & 1.03 (0.18) & NA & NA & 0.00 (0.00) \\ 
  \cellcolor{green!30}{MAGNA++ (PE-P-P)} & 1.27 (0.91) & 1.86e-01 & NA & 0.00 (0.00) \\ 
  \cellcolor{green!30}{GHOST (PE-P-P)} & 1.60 (1.16) & 1.31e-02 & 2.32e-01 & 0.00 (0.00) \\ 
  \cellcolor{green!30}{WAVE (PE-P-P)} & 1.60 (1.45) & 2.61e-02 & 9.42e-02 & 0.00 (0.00) \\ 
  \cellcolor{green!30}{LGRAAL (PE-P-P)} & 3.70 (2.00) & 3.77e-05 & 3.69e-05 & 0.00 (0.00) \\ 
  \cellcolor{blue!30}{multiMAGNA++ (PE-M-M)} & 4.97 (1.87) & 1.80e-06 & 2.84e-06 & 0.00 (0.00) \\ 
  \cellcolor{blue!30}{IsoRankN (PE-M-M)} & 7.13 (1.85) & 9.23e-07 & 9.36e-07 & 0.00 (0.00) \\ 
  \cellcolor{red!30}{IsoRankN (PE-M-P)} & 7.83 (2.44) & 1.47e-06 & 1.49e-06 & 0.00 (0.00) \\ 
  \cellcolor{blue!30}{ConvexAlign (PE-M-M)} & 8.37 (2.97) & 1.71e-06 & 2.03e-06 & 0.00 (0.00) \\ 
  \cellcolor{blue!30}{BEAMS (PE-M-M)} & 8.53 (3.50) & 5.01e-06 & 5.24e-06 & 0.00 (0.00) \\ 
  \cellcolor{red!30}{BEAMS (PE-M-P)} & 8.60 (3.56) & 5.24e-06 & 5.36e-06 & 0.00 (0.00) \\ 
  \cellcolor{red!30}{ConvexAlign (PE-M-P)} & 10.60 (1.81) & 5.04e-07 & 5.22e-07 & 0.00 (0.00) \\ 
   \bottomrule

\end{tabular}
\vspace{0.15cm}
\caption{Overall ranking of the NA methods for the {\bf PE framework}
over all evaluation tests (where a test is a combination of an NA
method, a network pair, and an alignment quality
measure) that use network pairs with \textbf{known node mapping}, for
T+S alignments, for both TQ and FQ measures. By NA
method, here, we mean the combination of a PNA or MNA method and the
alignment category (Section \ref{sec:evaluation} of the main paper). Namely, there are
12 NA methods in the PE framework (four PNA methods associated with
the PE-P-P categories and four MNA methods associated with each of the
PE-M-M and PE-M-P categories).
The table can be interpreted the same way as Supplementary Table
\ref{tab:exp2_ranktable_ts_all_topo}.} 
\label{tab:exp2_ranktable_ts_known_all}
\end{table}

\begin{table}[h!]
\centering
\begin{tabular}{rrrrr}
  \toprule
NA method & Overall rank & $p_1$-value & $p_2$-value & Non-sig (fail) \\ 
  \midrule
\cellcolor{red!30}{ConvexAlign (PE-M-P)} & 4.57 (3.66) & NA & NA & 0.06 (0.00) \\ 
  \cellcolor{green!30}{MAGNA++ (PE-P-P)} & 5.49 (2.64) & 4.81e-02 & NA & 0.09 (0.00) \\ 
  \cellcolor{red!30}{multiMAGNA++ (PE-M-P)} & 5.74 (3.84) & 3.70e-02 & 4.12e-01 & 0.16 (0.00) \\ 
  \cellcolor{green!30}{WAVE (PE-P-P)} & 6.23 (4.32) & 1.15e-02 & 1.52e-01 & 0.20 (0.00) \\ 
  \cellcolor{blue!30}{ConvexAlign (PE-M-M)} & 6.40 (4.31) & 9.33e-08 & 8.63e-02 & 0.17 (0.00) \\ 
  \cellcolor{green!30}{LGRAAL (PE-P-P)} & 6.71 (3.62) & 2.85e-03 & 1.66e-01 & 0.27 (0.07) \\ 
  \cellcolor{blue!30}{multiMAGNA++ (PE-M-M)} & 7.14 (3.89) & 3.71e-03 & 1.36e-04 & 0.29 (0.00) \\ 
  \cellcolor{green!30}{GHOST (PE-P-P)} & 7.99 (3.83) & 1.08e-05 & 1.83e-05 & 0.39 (0.21) \\ 
  \cellcolor{red!30}{BEAMS (PE-M-P)} & 8.47 (3.47) & 3.04e-08 & 2.33e-06 & 0.33 (0.00) \\ 
  \cellcolor{red!30}{IsoRankN (PE-M-P)} & 8.89 (3.69) & 1.43e-10 & 1.02e-07 & 0.46 (0.00) \\ 
  \cellcolor{blue!30}{IsoRankN (PE-M-M)} & 8.97 (3.46) & 4.28e-11 & 3.50e-07 & 0.43 (0.00) \\ 
  \cellcolor{blue!30}{BEAMS (PE-M-M)} & 9.13 (3.38) & 1.58e-09 & 7.46e-08 & 0.44 (0.00) \\ 
   \bottomrule

\end{tabular}
\vspace{0.15cm}
\caption{Overall ranking of the NA methods for the {\bf PE framework}
over all evaluation tests (where a test is a combination of an NA
method, a network pair, and an alignment quality
measure) that use network pairs with \textbf{unknown node mapping}, for
T+S alignments, for both TQ and FQ measures. By NA
method, here, we mean the combination of a PNA or MNA method and the
alignment category (Section \ref{sec:evaluation} of the main
paper). Namely, there are 
12 NA methods in the PE framework (four PNA methods associated with
the PE-P-P categories and four MNA methods associated with each of the
PE-M-M and PE-M-P categories).
The table can be interpreted the same way as Supplementary Table \ref{tab:exp2_ranktable_ts_all_topo}.}
\label{tab:exp2_ranktable_ts_unknown_all}
\end{table}

\begin{table}[h!]
\centering
\begin{tabular}{rrrrr}
  \toprule
NA method & Overall rank & $p_1$-value & $p_2$-value & Non-sig (fail) \\ 
  \midrule
\cellcolor{red!30}{multiMAGNA++ (ME-M-P)} & 1.71 (1.25) & NA & NA & 0.00 (0.00) \\ 
  \cellcolor{green!30}{WAVE (ME-P-P)} & 2.14 (1.46) & 3.56e-01 & NA & 0.00 (0.00) \\ 
  \cellcolor{green!30}{MAGNA++ (ME-P-P)} & 3.00 (2.31) & 1.01e-01 & 2.05e-01 & 0.00 (0.00) \\ 
  \cellcolor{green!30}{GHOST (ME-P-P)} & 3.86 (4.02) & 9.87e-02 & 1.40e-01 & 0.14 (0.00) \\ 
  \cellcolor{blue!30}{multiMAGNA++ (ME-M-M)} & 4.00 (2.00) & 4.46e-02 & 7.47e-02 & 0.00 (0.00) \\ 
  \cellcolor{green!30}{LGRAAL (ME-P-P)} & 5.00 (3.70) & 3.66e-02 & 2.90e-02 & 0.14 (0.00) \\ 
  \cellcolor{blue!30}{IsoRankN (ME-M-M)} & 7.57 (1.13) & 1.08e-02 & 1.11e-02 & 0.00 (0.00) \\ 
  \cellcolor{blue!30}{ConvexAlign (ME-M-M)} & 8.71 (2.36) & 1.08e-02 & 1.07e-02 & 0.00 (0.00) \\ 
  \cellcolor{blue!30}{BEAMS (ME-M-M)} & 9.14 (2.54) & 1.10e-02 & 1.10e-02 & 0.29 (0.00) \\ 
  \cellcolor{red!30}{IsoRankN (ME-M-P)} & 10.43 (1.99) & 1.09e-02 & 1.10e-02 & 0.57 (0.00) \\ 
  \cellcolor{red!30}{BEAMS (ME-M-P)} & 10.71 (1.89) & 1.07e-02 & 1.09e-02 & 0.57 (0.00) \\ 
  \cellcolor{red!30}{ConvexAlign (ME-M-P)} & 11.43 (0.98) & 9.95e-03 & 1.05e-02 & 0.43 (0.00) \\ 
   \bottomrule

\end{tabular}
\vspace{0.15cm}
\caption{Overall ranking of the NA methods for the {\bf ME framework}
over all evaluation tests (where a test is a combination of an NA
method, a network set, and an alignment quality measure) that use
{\bf TQ measures}, for T+S alignments, for networks with both known
and unknown node mapping.    By NA
method, here, we mean the combination of a PNA or MNA method and the
alignment category (Section \ref{sec:evaluation} of the main paper). Namely, there are
12 NA methods in the ME framework (four PNA methods associated with
the ME-P-P categories and four MNA methods associated with each of the
ME-M-M and ME-M-P categories).
The table can be interpreted the same way as Supplementary Table \ref{tab:exp2_ranktable_ts_all_topo}.}
\label{tab:exp1_ranktable_ts_all_topo}
\end{table}

\begin{table}[h!]
\centering
\begin{tabular}{rrrrr}
  \toprule
NA method & Overall rank & $p_1$-value & $p_2$-value & Non-sig (fail) \\ 
  \midrule
\cellcolor{green!30}{MAGNA++ (ME-P-P)} & 4.22 (2.82) & NA & NA & 0.00 (0.00) \\ 
  \cellcolor{blue!30}{ConvexAlign (ME-M-M)} & 5.11 (3.82) & 3.83e-01 & NA & 0.00 (0.00) \\ 
  \cellcolor{red!30}{ConvexAlign (ME-M-P)} & 5.44 (5.27) & 3.83e-01 & 6.12e-01 & 0.11 (0.00) \\ 
  \cellcolor{green!30}{LGRAAL (ME-P-P)} & 5.78 (4.18) & 3.67e-01 & 3.63e-01 & 0.22 (0.00) \\ 
  \cellcolor{green!30}{GHOST (ME-P-P)} & 5.89 (4.59) & 1.75e-01 & 4.06e-01 & 0.11 (0.00) \\ 
  \cellcolor{red!30}{multiMAGNA++ (ME-M-P)} & 5.89 (3.98) & 7.13e-02 & 4.06e-01 & 0.11 (0.00) \\ 
  \cellcolor{blue!30}{IsoRankN (ME-M-M)} & 6.00 (4.00) & 2.20e-01 & 2.19e-01 & 0.22 (0.00) \\ 
  \cellcolor{green!30}{WAVE (ME-P-P)} & 7.00 (4.47) & 1.07e-02 & 2.36e-01 & 0.11 (0.00) \\ 
  \cellcolor{blue!30}{multiMAGNA++ (ME-M-M)} & 7.33 (4.18) & 2.36e-02 & 2.38e-01 & 0.11 (0.00) \\ 
  \cellcolor{blue!30}{BEAMS (ME-M-M)} & 7.56 (4.67) & 5.32e-02 & 9.04e-02 & 0.33 (0.00) \\ 
  \cellcolor{red!30}{IsoRankN (ME-M-P)} & 8.78 (3.90) & 2.09e-02 & 5.98e-02 & 0.44 (0.00) \\ 
  \cellcolor{red!30}{BEAMS (ME-M-P)} & 9.00 (4.39) & 2.10e-02 & 4.13e-02 & 0.56 (0.00) \\ 
   \bottomrule

\end{tabular}
\vspace{0.15cm}
\caption{Overall ranking of the NA methods for the {\bf ME framework}
over all evaluation tests (where a test is a combination of an NA
method, a network set, and an alignment quality measure) that use
{\bf FQ measures}, for T+S alignments, for networks with both known
and unknown node mapping.   By NA
method, here, we mean the combination of a PNA or MNA method and the
alignment category (Section \ref{sec:evaluation} of the main paper). Namely, there are
12 NA methods in the ME framework (four PNA methods associated with
the ME-P-P categories and four MNA methods associated with each of the
ME-M-M and ME-M-P categories).
The table can be interpreted the same way as Supplementary Table \ref{tab:exp2_ranktable_ts_all_topo}.}
\label{tab:exp1_ranktable_ts_all_func}
\end{table}

\begin{table}[h!]
\centering
\begin{tabular}{rrrrr}
  \toprule
NA method & Overall rank & $p_1$-value & $p_2$-value & Non-sig (fail) \\ 
  \midrule
\cellcolor{green!30}{GHOST (ME-P-P)} & 1.17 (0.41) & NA & NA & 0.00 (0.00) \\ 
  \cellcolor{red!30}{multiMAGNA++ (ME-M-P)} & 1.33 (0.82) & 5.00e-01 & NA & 0.00 (0.00) \\ 
  \cellcolor{green!30}{MAGNA++ (ME-P-P)} & 1.50 (1.22) & 5.00e-01 & 5.00e-01 & 0.00 (0.00) \\ 
  \cellcolor{green!30}{WAVE (ME-P-P)} & 2.17 (1.83) & 1.73e-01 & 1.86e-01 & 0.00 (0.00) \\ 
  \cellcolor{green!30}{LGRAAL (ME-P-P)} & 3.17 (2.40) & 7.45e-02 & 8.68e-02 & 0.00 (0.00) \\ 
  \cellcolor{blue!30}{multiMAGNA++ (ME-M-M)} & 4.17 (2.48) & 4.96e-02 & 4.96e-02 & 0.00 (0.00) \\ 
  \cellcolor{blue!30}{IsoRankN (ME-M-M)} & 6.33 (2.66) & 2.39e-02 & 2.72e-02 & 0.00 (0.00) \\ 
  \cellcolor{red!30}{IsoRankN (ME-M-P)} & 7.33 (3.20) & 2.67e-02 & 2.39e-02 & 0.00 (0.00) \\ 
  \cellcolor{blue!30}{BEAMS (ME-M-M)} & 8.17 (3.60) & 2.67e-02 & 2.39e-02 & 0.00 (0.00) \\ 
  \cellcolor{red!30}{BEAMS (ME-M-P)} & 8.50 (4.04) & 2.84e-02 & 2.72e-02 & 0.17 (0.00) \\ 
  \cellcolor{blue!30}{ConvexAlign (ME-M-M)} & 10.33 (1.03) & 1.55e-02 & 1.55e-02 & 0.00 (0.00) \\ 
  \cellcolor{red!30}{ConvexAlign (ME-M-P)} & 11.17 (2.04) & 1.31e-02 & 1.31e-02 & 0.17 (0.00) \\ 
   \bottomrule

\end{tabular}
\vspace{0.15cm}
\caption{Overall ranking of the NA methods for the {\bf ME framework}
over all evaluation tests (where a test is a combination of an NA
method, a network set, and an alignment quality
measure) that use network pairs with \textbf{known node mapping}, for
T+S alignments, for both TQ and FQ measures. By NA
method, here, we mean the combination of a PNA or MNA method and the
alignment category (Section \ref{sec:evaluation}). Namely, there are
12 NA methods in the ME framework (four PNA methods associated with
the ME-P-P categories and four MNA methods associated with each of the
ME-M-M and ME-M-P categories).
The table can be interpreted the same way as Supplementary Table \ref{tab:exp2_ranktable_ts_all_topo}.}
\label{tab:exp1_ranktable_ts_known_all}
\end{table}

\begin{table}[h!]
\centering
\begin{tabular}{rrrrr}
  \toprule
NA method & Overall rank & $p_1$-value & $p_2$-value & Non-sig (fail) \\ 
  \midrule
\cellcolor{blue!30}{ConvexAlign (ME-M-M)} & 4.50 (2.76) & NA & NA & 0.00 (0.00) \\ 
  \cellcolor{green!30}{MAGNA++ (ME-P-P)} & 5.00 (2.31) & 3.60e-01 & NA & 0.00 (0.00) \\ 
  \cellcolor{red!30}{multiMAGNA++ (ME-M-P)} & 5.70 (3.80) & 2.21e-01 & 4.39e-01 & 0.10 (0.00) \\ 
  \cellcolor{red!30}{ConvexAlign (ME-M-P)} & 6.20 (5.33) & 1.42e-01 & 3.04e-01 & 0.30 (0.00) \\ 
  \cellcolor{green!30}{WAVE (ME-P-P)} & 6.50 (4.45) & 1.92e-01 & 2.06e-01 & 0.10 (0.00) \\ 
  \cellcolor{green!30}{LGRAAL (ME-P-P)} & 6.80 (4.02) & 1.06e-01 & 3.41e-01 & 0.30 (0.00) \\ 
  \cellcolor{blue!30}{IsoRankN (ME-M-M)} & 6.90 (3.48) & 3.96e-03 & 1.92e-01 & 0.20 (0.00) \\ 
  \cellcolor{blue!30}{multiMAGNA++ (ME-M-M)} & 6.90 (4.07) & 1.10e-01 & 1.00e-01 & 0.10 (0.00) \\ 
  \cellcolor{green!30}{GHOST (ME-P-P)} & 7.30 (3.95) & 6.28e-02 & 1.20e-01 & 0.20 (0.00) \\ 
  \cellcolor{blue!30}{BEAMS (ME-M-M)} & 8.30 (4.19) & 2.05e-02 & 6.30e-02 & 0.50 (0.00) \\ 
  \cellcolor{red!30}{BEAMS (ME-M-P)} & 10.50 (3.17) & 2.86e-03 & 7.06e-03 & 0.80 (0.00) \\ 
  \cellcolor{red!30}{IsoRankN (ME-M-P)} & 10.80 (2.57) & 2.82e-03 & 7.06e-03 & 0.80 (0.00) \\ 
   \bottomrule

\end{tabular}
\vspace{0.15cm}
\caption{Overall ranking of the NA methods for the {\bf ME framework}
over all evaluation tests (where a test is a combination of an NA
method, a network set, and an alignment quality
measure) that use network pairs with \textbf{unknown node mapping}, for
T+S alignments, for both TQ and FQ measures. By NA
method, here, we mean the combination of a PNA or MNA method and the
alignment category (Section \ref{sec:evaluation} of the main paper). Namely, there are
12 NA methods in the ME framework (four PNA methods associated with
the ME-P-P categories and four MNA methods associated with each of the
ME-M-M and ME-M-P categories).
The table can be interpreted the same way as Supplementary Table \ref{tab:exp2_ranktable_ts_all_topo}.}
\label{tab:exp1_ranktable_ts_unknown_all}
\end{table}

\begin{table}[h!]
\centering
\begin{tabular}{rrrrr}
  \toprule
NA method & Overall rank  \\ 
  \midrule
  \cellcolor{blue!30}{multiMAGNA++ (ME-M-M)} & 2.25 (1.50)  \\ 
  \cellcolor{green!30}{MAGNA++ (ME-P-P)} & 3.25 (3.20)  \\ 
  \cellcolor{red!30}{ConvexAlign (ME-M-P)} & 4.25 (1.71)  \\ 
  \cellcolor{green!30}{GHOST (ME-P-P)} & 4.25 (2.36)  \\ 
  \cellcolor{green!30}{LGRAAL (ME-P-P)} & 4.25 (1.50)  \\ 
  \cellcolor{green!30}{WAVE (ME-P-P)} & 5.00 (2.45)  \\ 
  \cellcolor{red!30}{multiMAGNA++ (ME-M-P)} & 6.25 (1.50)  \\ 
  \cellcolor{blue!30}{IsoRankN (ME-M-M)} & 7.75 (3.30)  \\ 
  \cellcolor{blue!30}{ConvexAlign (ME-M-M)} & 8.25 (0.96) \\ 
  \cellcolor{red!30}{IsoRankN (ME-M-P)} & 9.50 (0.58)  \\ 
  \bottomrule
\end{tabular}
\vspace{0.15cm}
\caption{\textcolor{black}{Overall ranking of the NA methods for the {\bf ME framework}
over all evaluation tests (where a test is a combination of an NA
method and a network set) that use the \textbf{mean normalized entropy measure}, for T alignments. By NA
method, here, we mean the combination of a PNA or MNA method and the
alignment category (Section \ref{sec:evaluation} of the main paper). Namely, there are
12 NA methods in the ME framework (four PNA methods associated with
the ME-P-P categories and four MNA methods associated with each of the
ME-M-M and ME-M-P categories). The alignment categories are color
coded. The ``Overall rank'' column shows the rank of each method averaged over all evaluation tests, along with the
corresponding standard deviation (in brackets). Since there are 12 methods in a given framework, the possible ranks
range from 1 to 12. The lower the rank, the better the given method.}}
\label{tab:entropy_t}
\end{table}

\begin{table}[h!]
\centering
\begin{tabular}{rrrrr}
  \toprule
NA method & Overall rank  \\ 
  \midrule
  \cellcolor{green!30}{LGRAAL (ME-P-P)} & 3.5 (1.00)  \\ 
  \cellcolor{blue!30}{IsoRankN (ME-M-M)} & 4.25 (4.72)  \\ 
  \cellcolor{red!30}{multiMAGNA++ (ME-M-P)} & 5.25 (4.35)  \\ 
  \cellcolor{green!30}{MAGNA++ (ME-P-P)} & 5.50 (4.43)  \\ 
  \cellcolor{blue!30}{ConvexAlign (ME-M-M)} & 6.75 (2.22) \\ 
  \cellcolor{blue!30}{multiMAGNA++ (ME-M-M)} & 7.00 (4.05)  \\ 
  \cellcolor{green!30}{WAVE (ME-P-P)} & 7.00 (2.94)  \\ 
  \cellcolor{red!30}{BEAMS (ME-M-P)} & 7.25 (4.11)  \\ 
  \cellcolor{red!30}{IsoRankN (ME-M-P)} & 7.5 (4.20)  \\ 
  \cellcolor{green!30}{GHOST (ME-P-P)} & 7.5 (4.79)  \\ 
  \cellcolor{blue!30}{BEAMS (ME-M-M)} & 8.25 (1.50)  \\ 
  \cellcolor{red!30}{ConvexAlign (ME-M-P)} & 8.25 (2.06)  \\

  \bottomrule
\end{tabular}
\vspace{0.15cm}
\caption{\textcolor{black}{Overall ranking of the NA methods for the {\bf ME framework}
over all evaluation tests (where a test is a combination of an NA
method and a network set) that use the \textbf{mean normalized entropy measure}, for T+S alignments. By NA
method, here, we mean the combination of a PNA or MNA method and the
alignment category (Section \ref{sec:evaluation} of the main paper). Namely, there are
12 NA methods in the ME framework (four PNA methods associated with
the ME-P-P categories and four MNA methods associated with each of the
ME-M-M and ME-M-P categories).
The table can be interpreted the same way as Supplementary Table \ref{tab:entropy_t}.}}
\label{tab:entropy_ts}
\end{table}

\begin{table}[h!]
  \centering
  \url{https://nd.edu/~cone/PNA_MNA/table_pef.csv}
\caption{Detailed alignment quality scores for the {\bf PE framework}.}
\label{tab:scores2}
\end{table}

\begin{table}[h!]
  \centering
 \url{https://nd.edu/~cone/PNA_MNA/table_mef.csv}
\caption{Detailed alignment quality scores for the {\bf ME framework}.}
\label{tab:scores1}
\end{table}

\clearpage
\newpage

\subsection*{Method rankings including GEDEVO-M}

\begin{table}[ht!]
\centering
\begin{tabular}{rrrrr}
  \toprule
NA method & Overall rank & $p_1$-value & $p_2$-value & Non-sig (fail) \\ 
  \midrule
\cellcolor{red!30}{multiMAGNA++ (ME-M-P)} & 1.71 (1.25) & NA & NA & 0.00 (0.00) \\ 
  \cellcolor{green!30}{WAVE (ME-P-P)} & 2.29 (1.60) & 2.85e-01 & NA & 0.00 (0.00) \\ 
  \cellcolor{green!30}{MAGNA++ (ME-P-P)} & 3.29 (2.75) & 9.87e-02 & 2.05e-01 & 0.00 (0.00) \\ 
  \cellcolor{green!30}{GHOST (ME-P-P)} & 4.00 (4.08) & 1.01e-01 & 1.40e-01 & 0.14 (0.00) \\ 
  \cellcolor{blue!30}{multiMAGNA++ (ME-M-M)} & 4.14 (1.86) & 3.67e-02 & 7.47e-02 & 0.00 (0.00) \\ 
  \cellcolor{green!30}{LGRAAL (ME-P-P)} & 5.14 (3.63) & 2.92e-02 & 2.90e-02 & 0.14 (0.00) \\ 
  \cellcolor{blue!30}{IsoRankN (ME-M-M)} & 7.86 (1.35) & 1.07e-02 & 1.12e-02 & 0.00 (0.00) \\ 
  \cellcolor{blue!30}{GEDEVO-M (ME-M-M)} & 8.80 (4.66) & 2.95e-02 & 5.28e-02 & 0.00 (0.00) \\ 
  \cellcolor{blue!30}{ConvexAlign (ME-M-M)} & 9.14 (2.41) & 1.10e-02 & 1.11e-02 & 0.00 (0.00) \\ 
  \cellcolor{blue!30}{BEAMS (ME-M-M)} & 9.43 (2.64) & 1.09e-02 & 1.10e-02 & 0.29 (0.00) \\ 
  \cellcolor{red!30}{IsoRankN (ME-M-P)} & 10.71 (2.29) & 1.10e-02 & 1.11e-02 & 0.57 (0.00) \\ 
  \cellcolor{red!30}{BEAMS (ME-M-P)} & 11.00 (2.16) & 1.01e-02 & 1.10e-02 & 0.57 (0.00) \\ 
  \cellcolor{red!30}{ConvexAlign (ME-M-P)} & 12.00 (1.15) & 1.07e-02 & 1.08e-02 & 0.43 (0.00) \\ 
   \bottomrule

\end{tabular}
\vspace{0.15cm}
\caption{Overall ranking of the NA methods for the {\bf ME framework} over all evaluation tests (where a test is a combination of an NA method, a network set, and an alignment quality measure) that use \textbf{TQ measures}, for T+S alignments, for networks with both known and unknown node mapping. The table mimics the analyses from Supplementary Table \ref{tab:exp1_ranktable_ts_all_topo} with the inclusion of an additional method, GEDEVO-M associated with the ME-M-M category.}
\label{tab:exp1_gedevom_ts_all_topo}
\end{table}

\begin{table}[ht!]
\centering
\begin{tabular}{rrrrr}
  \toprule
NA method & Overall rank & $p_1$-value & $p_2$-value & Non-sig (fail) \\ 
  \midrule
\cellcolor{green!30}{MAGNA++ (ME-P-P)} & 4.22 (2.82) & NA & NA & 0.00 (0.00) \\ 
  \cellcolor{blue!30}{ConvexAlign (ME-M-M)} & 5.11 (3.82) & 3.83e-01 & NA & 0.00 (0.00) \\ 
  \cellcolor{red!30}{ConvexAlign (ME-M-P)} & 5.56 (5.43) & 3.83e-01 & 6.12e-01 & 0.11 (0.00) \\ 
  \cellcolor{green!30}{LGRAAL (ME-P-P)} & 5.89 (4.37) & 3.67e-01 & 3.37e-01 & 0.22 (0.00) \\ 
  \cellcolor{green!30}{GHOST (ME-P-P)} & 6.00 (4.77) & 1.75e-01 & 3.83e-01 & 0.11 (0.00) \\ 
  \cellcolor{red!30}{multiMAGNA++ (ME-M-P)} & 6.00 (4.18) & 7.13e-02 & 3.83e-01 & 0.11 (0.00) \\ 
  \cellcolor{blue!30}{IsoRankN (ME-M-M)} & 6.11 (4.20) & 2.20e-01 & 2.02e-01 & 0.22 (0.00) \\ 
  \cellcolor{green!30}{WAVE (ME-P-P)} & 7.11 (4.62) & 1.07e-02 & 2.20e-01 & 0.11 (0.00) \\ 
  \cellcolor{blue!30}{multiMAGNA++ (ME-M-M)} & 7.44 (4.33) & 2.36e-02 & 2.20e-01 & 0.11 (0.00) \\ 
  \cellcolor{blue!30}{BEAMS (ME-M-M)} & 7.67 (4.80) & 5.32e-02 & 8.02e-02 & 0.33 (0.00) \\ 
  \cellcolor{red!30}{IsoRankN (ME-M-P)} & 9.00 (4.12) & 2.07e-02 & 4.80e-02 & 0.44 (0.00) \\ 
  \cellcolor{red!30}{BEAMS (ME-M-P)} & 9.33 (4.66) & 2.10e-02 & 3.28e-02 & 0.56 (0.00) \\ 
  \cellcolor{blue!30}{GEDEVO-M (ME-M-M)} & 12.50 (0.84) & 1.68e-02 & 1.78e-02 & 0.33 (0.00) \\ 
   \bottomrule

\end{tabular}
\vspace{0.15cm}
\caption{Overall ranking of the NA methods for the {\bf ME framework} over all evaluation tests (where a test is a combination of an NA method, a network set, and an alignment quality measure) that use \textbf{FQ measures}, for T+S alignments, for networks with both known and unknown node mapping. The table mimics the analyses from Supplementary Table \ref{tab:exp1_ranktable_ts_all_func} with the inclusion of an additional method, GEDEVO-M associated with the ME-M-M category.}

\label{tab:exp1_gedevom_ts_all_func}
\end{table}

\begin{table}[ht!]
\centering
\begin{tabular}{rrrrr}
  \toprule
NA method & Overall rank & $p_1$-value & $p_2$-value & Non-sig (fail) \\ 
  \midrule
\cellcolor{green!30}{GHOST (ME-P-P)} & 1.17 (0.41) & NA & NA & 0.00 (0.00) \\ 
  \cellcolor{red!30}{multiMAGNA++ (ME-M-P)} & 1.33 (0.82) & 5.00e-01 & NA & 0.00 (0.00) \\ 
  \cellcolor{green!30}{MAGNA++ (ME-P-P)} & 1.50 (1.22) & 5.00e-01 & 5.00e-01 & 0.00 (0.00) \\ 
  \cellcolor{green!30}{WAVE (ME-P-P)} & 2.17 (1.83) & 1.73e-01 & 1.86e-01 & 0.00 (0.00) \\ 
  \cellcolor{green!30}{LGRAAL (ME-P-P)} & 3.17 (2.40) & 7.45e-02 & 8.68e-02 & 0.00 (0.00) \\ 
  \cellcolor{blue!30}{multiMAGNA++ (ME-M-M)} & 4.17 (2.48) & 4.96e-02 & 4.96e-02 & 0.00 (0.00) \\ 
  \cellcolor{blue!30}{IsoRankN (ME-M-M)} & 6.33 (2.66) & 2.39e-02 & 2.72e-02 & 0.00 (0.00) \\ 
  \cellcolor{red!30}{IsoRankN (ME-M-P)} & 7.33 (3.20) & 2.67e-02 & 2.39e-02 & 0.00 (0.00) \\ 
  \cellcolor{blue!30}{BEAMS (ME-M-M)} & 8.17 (3.60) & 2.67e-02 & 2.39e-02 & 0.00 (0.00) \\ 
  \cellcolor{red!30}{BEAMS (ME-M-P)} & 8.67 (4.23) & 2.90e-02 & 2.84e-02 & 0.17 (0.00) \\ 
  \cellcolor{blue!30}{ConvexAlign (ME-M-M)} & 10.50 (1.22) & 1.70e-02 & 1.70e-02 & 0.00 (0.00) \\ 
  \cellcolor{red!30}{ConvexAlign (ME-M-P)} & 11.50 (2.26) & 1.68e-02 & 1.68e-02 & 0.17 (0.00) \\ 
  \cellcolor{blue!30}{GEDEVO-M (ME-M-M)} & 12.33 (0.82) & 1.68e-02 & 1.70e-02 & 0.00 (0.00) \\ 
   \bottomrule

\end{tabular}
\vspace{0.15cm}
\caption{Overall ranking of the NA methods for the {\bf ME framework} over all evaluation tests (where a test is a combination of an NA method, a network set, and an alignment quality measure) that use network pairs with \textbf{known node mapping}, for T+S alignments, for both TQ and FQ measures. The table mimics the analyses from Supplementary Table \ref{tab:exp1_ranktable_ts_known_all} with the inclusion of an additional method, GEDEVO-M associated with the ME-M-M category.}

\label{tab:exp1_gedevom_ts_known_all}
\end{table}

\begin{table}[ht!]
\centering
\begin{tabular}{rrrrr}
  \toprule
NA method & Overall rank & $p_1$-value & $p_2$-value & Non-sig (fail) \\ 
  \midrule
\cellcolor{blue!30}{ConvexAlign (ME-M-M)} & 4.70 (3.02) & NA & NA & 0.00 (0.00) \\ 
  \cellcolor{green!30}{MAGNA++ (ME-P-P)} & 5.20 (2.44) & 3.60e-01 & NA & 0.00 (0.00) \\ 
  \cellcolor{red!30}{multiMAGNA++ (ME-M-P)} & 5.80 (3.99) & 2.86e-01 & 4.80e-01 & 0.10 (0.00) \\ 
  \cellcolor{red!30}{ConvexAlign (ME-M-P)} & 6.50 (5.66) & 1.42e-01 & 3.04e-01 & 0.30 (0.00) \\ 
  \cellcolor{green!30}{WAVE (ME-P-P)} & 6.70 (4.52) & 2.07e-01 & 2.06e-01 & 0.10 (0.00) \\ 
  \cellcolor{green!30}{LGRAAL (ME-P-P)} & 7.00 (4.08) & 1.17e-01 & 3.41e-01 & 0.30 (0.00) \\ 
  \cellcolor{blue!30}{multiMAGNA++ (ME-M-M)} & 7.10 (4.09) & 1.10e-01 & 1.10e-01 & 0.10 (0.00) \\ 
  \cellcolor{blue!30}{IsoRankN (ME-M-M)} & 7.20 (3.74) & 3.96e-03 & 1.92e-01 & 0.20 (0.00) \\ 
  \cellcolor{green!30}{GHOST (ME-P-P)} & 7.50 (4.03) & 7.61e-02 & 1.30e-01 & 0.20 (0.00) \\ 
  \cellcolor{blue!30}{BEAMS (ME-M-M)} & 8.60 (4.38) & 2.06e-02 & 6.30e-02 & 0.50 (0.00) \\ 
  \cellcolor{blue!30}{GEDEVO-M (ME-M-M)} & 9.00 (4.85) & 1.39e-01 & 2.05e-01 & 0.40 (0.00) \\ 
  \cellcolor{red!30}{BEAMS (ME-M-P)} & 10.90 (3.41) & 2.91e-03 & 7.12e-03 & 0.80 (0.00) \\ 
  \cellcolor{red!30}{IsoRankN (ME-M-P)} & 11.20 (2.82) & 2.86e-03 & 7.12e-03 & 0.80 (0.00) \\ 
   \bottomrule

\end{tabular}
\vspace{0.15cm}
\caption{Overall ranking of the NA methods for the {\bf ME framework} over all evaluation tests (where a test is a combination of an NA method, a network set, and an alignment quality measure) that use networks pairs with \textbf{unknown node mapping}, for T+S alignments, for both TQ and FQ measures. The table mimics the analyses from Supplementary Table \ref{tab:exp1_ranktable_ts_unknown_all} with the inclusion of an additional method, GEDEVO-M associated with the ME-M-M category.}

\label{tab:exp1_gedevom_ts_unknown_all}
\end{table}

\begin{table}[ht!]
\centering
\begin{tabular}{rrrrr}
  \toprule
NA method & Overall rank & $p_1$-value & $p_2$-value & Non-sig (fail) \\ 
  \midrule
\cellcolor{green!30}{MAGNA++ (ME-P-P)} & 3.81 (2.74) & NA & NA & 0.00 (0.00) \\ 
  \cellcolor{red!30}{multiMAGNA++ (ME-M-P)} & 4.12 (3.84) & 5.18e-01 & NA & 0.06 (0.00) \\ 
  \cellcolor{green!30}{WAVE (ME-P-P)} & 5.00 (4.31) & 1.26e-01 & 3.91e-02 & 0.06 (0.00) \\ 
  \cellcolor{green!30}{GHOST (ME-P-P)} & 5.12 (4.46) & 1.98e-01 & 1.52e-01 & 0.12 (0.00) \\ 
  \cellcolor{green!30}{LGRAAL (ME-P-P)} & 5.56 (3.95) & 1.24e-01 & 8.38e-02 & 0.19 (0.00) \\ 
  \cellcolor{blue!30}{multiMAGNA++ (ME-M-M)} & 6.00 (3.78) & 1.87e-02 & 5.39e-03 & 0.06 (0.00) \\ 
  \cellcolor{blue!30}{ConvexAlign (ME-M-M)} & 6.88 (3.79) & 3.91e-02 & 8.88e-02 & 0.00 (0.00) \\ 
  \cellcolor{blue!30}{IsoRankN (ME-M-M)} & 6.88 (3.30) & 1.32e-02 & 1.97e-02 & 0.12 (0.00) \\ 
  \cellcolor{red!30}{ConvexAlign (ME-M-P)} & 8.38 (5.21) & 1.68e-02 & 1.42e-02 & 0.25 (0.00) \\ 
  \cellcolor{blue!30}{BEAMS (ME-M-M)} & 8.44 (3.98) & 3.42e-03 & 5.35e-03 & 0.31 (0.00) \\ 
  \cellcolor{red!30}{IsoRankN (ME-M-P)} & 9.75 (3.45) & 6.25e-04 & 1.01e-03 & 0.50 (0.00) \\ 
  \cellcolor{red!30}{BEAMS (ME-M-P)} & 10.06 (3.77) & 6.50e-04 & 1.21e-03 & 0.56 (0.00) \\ 
  \cellcolor{blue!30}{GEDEVO-M (ME-M-M)} & 10.82 (3.57) & 5.36e-03 & 2.90e-03 & 0.18 (0.00) \\ 
   \bottomrule

\end{tabular}
\vspace{0.15cm}
\caption{Overall ranking of the NA methods for the {\bf ME framework} over all evaluation tests (where a test is a combination of an NA method, a network set, and an alignment quality measure) for T+S alignments, for both TQ and FQ measures, for networks with both known and unknown node mapping. The table mimics the analyses from View I of Figure 5 from the main paper, with the inclusion of an additional method, GEDEVO-M associated with the ME-M-M category.}
\label{tab:exp1_gedevom_ts_all_all}
\end{table}

\clearpage
\newpage

\section*{SUPPLEMENTARY FIGURES}

\begin{figure}[h!]
\centering
\hspace*{-1em}\includegraphics[width=0.65\linewidth]{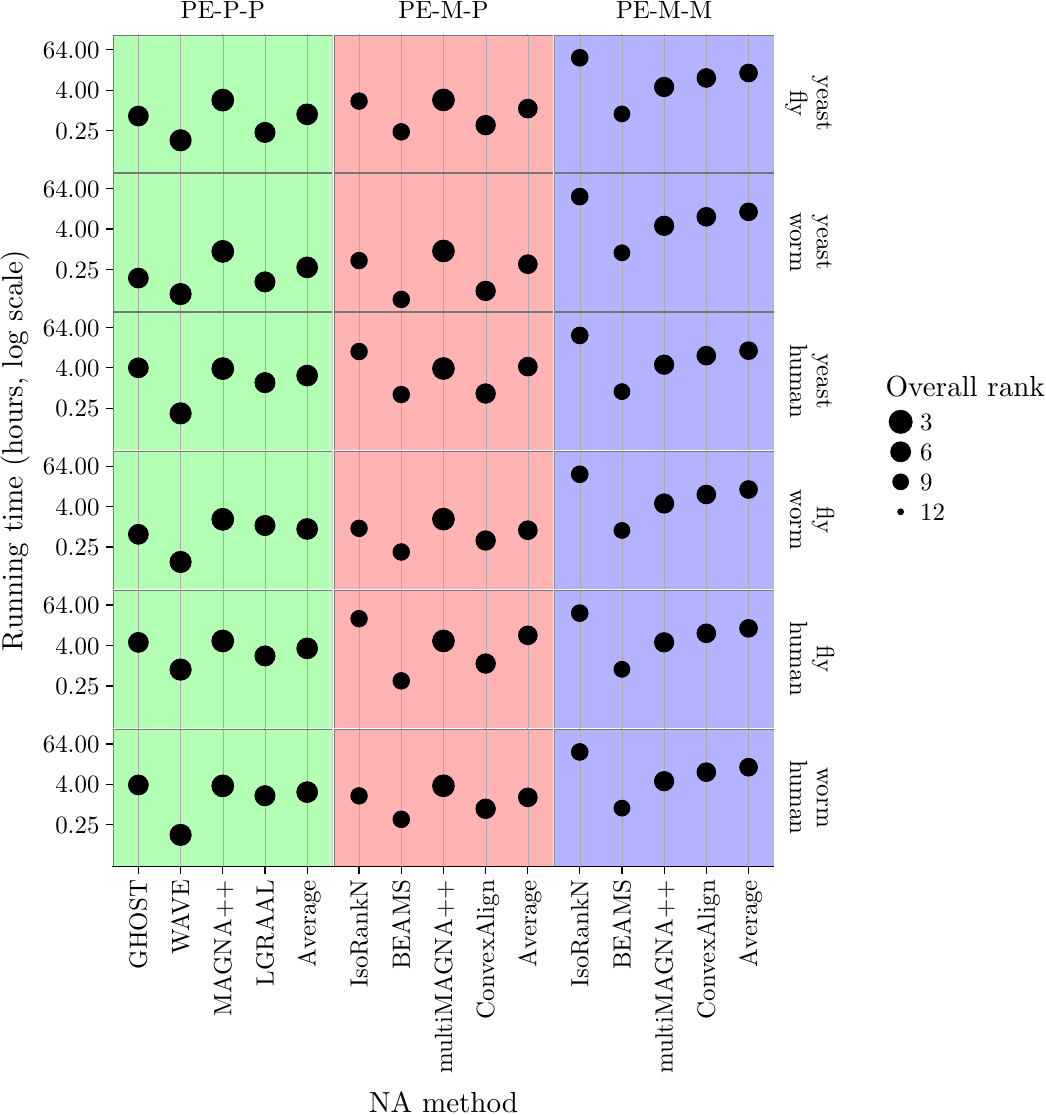}%
\caption{Overall ranking of an NA method versus its running time for the {\bf PE framework}
over all evaluation tests (where a test is a combination of an NA
method, a network pair, and an alignment quality measure).  By NA method, here, we
mean the combination of a PNA or MNA method and the alignment category
(Section \ref{sec:evaluation}). Namely, there are 12 NA methods in the
PE framework (four PNA methods associated with the PE-P-P categories
and four MNA methods associated with each of the PE-M-M and PE-M-P
categories). The alignment categories are color coded.
The running time results are when aligning all
network pairs in the Y2H$_1$ network set, where
each method is restricted to use a {\bf single core}. The size of
each point visualizes the overall ranking of the corresponding method
over all evaluation tests over all network pairs/sets, corresponding
to the ``Overall rank'' column in View I of Fig. \ref{fig:res} in
the main paper; the larger the point size,
the better the method. In order to allow for easier comparison between
the different alignment categories, ``Average'' shows the average
running times and average rankings of the methods in each alignment
category.}
\label{fig:exp5_pef}
\end{figure}

\begin{figure}[h!]
\centering
\hspace*{-1em}\includegraphics[width=0.5\linewidth]{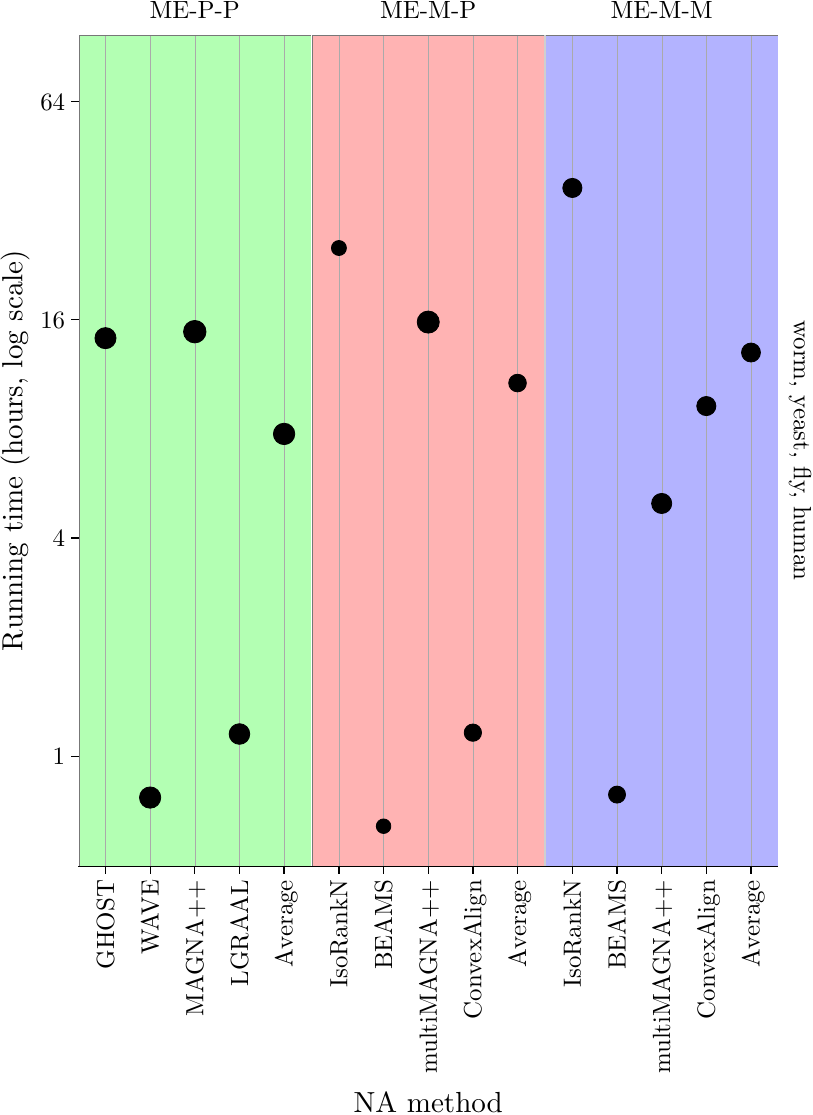}%
\caption{Overall ranking of an NA method versus its running time for the {\bf ME framework}
over all evaluation tests (where a test is a combination of an NA
method, a network pair, and an alignment quality measure).  By NA method, here, we
mean the combination of a PNA or MNA method and the alignment category
(Section \ref{sec:evaluation} of the main paper). Namely, there are 12 NA methods in the
ME framework (four PNA methods associated with the ME-P-P categories
and four MNA methods associated with each of the ME-M-M and ME-M-P
categories). The alignment categories are color coded.
The running time results are when  aligning the Y2H$_1$ network set, where
each method is restricted to use a {\bf single core}. The size of
each point visualizes the overall ranking of the corresponding method
over all evaluation tests over all network pairs/sets, corresponding
to the ``Overall rank'' column in View I of Fig. \ref{fig:res} in
the main paper; the larger the point size,
the better the method. In order to allow for easier comparison between
the different alignment categories, ``Average'' shows the average
running times and average rankings of the methods in each alignment
category.}
\label{fig:exp5_mef}
\end{figure}

\begin{figure*}
\centering
\includegraphics[width=0.9\linewidth]{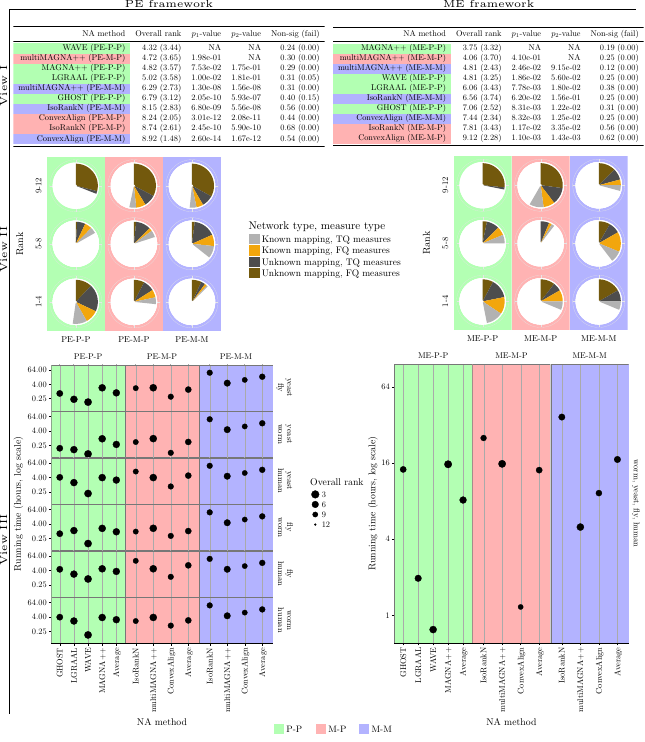}
\vspace{-0.3cm}
\caption{\tiny Method comparison results for each of the {\bf PE} and {\bf
ME} frameworks over all evaluation tests (where a test is a
combination of an NA method, a network pair/set, and an alignment
quality measure), for T alignments.  By NA method, here, we mean the combination of a
PNA or MNA method and the alignment category (Section
\ref{sec:evaluation}). Namely, there are 12 NA methods in the PE
framework (four PNA methods associated with the PE-P-P categories and
four MNA methods associated with each of the PE-M-M and PE-M-P
categories) and 12 NA methods in the ME framework (four PNA methods
associated with the ME-P-P categories and four MNA methods associated
with each of the ME-M-M and ME-M-P categories). The alignment
categories are color coded.  {\bf View I.} Overall ranking of the NA
methods. The ``Overall rank'' column shows the rank of each method
averaged over all evaluation tests, along with the corresponding
standard deviation (in brackets). Since there are 12 methods in a
given framework, the possible ranks range from 1 to 12. The lower the
rank, the better the given method.  The ``$p_1$-value'' column shows
the statistical significance of the difference between the ranking of
each method and the 1$^{st}$ best ranked method.  The ``$p_2$-value''
column shows the statistical significance of the difference between
the ranking of each method and the 2$^{nd}$ best ranked method.  The
``Non. sig. (fail)'' column shows the fraction of evaluation tests in
which the alignment quality score is not statistically significant,
and, in brackets, the fraction of evaluation tests in which the given
NA method failed to produce an alignment.  Equivalent results over all
evaluation tests broken down into functional and topological alignment
quality measures, as well as over all evaluation tests broken down
into network pairs/sets with known and unknown node mapping, are shown
in Supplementary Tables
\ref{tab:exp2_ranktable_ts_all_topo}--\ref{tab:exp1_ranktable_ts_unknown_all}.
{\bf View II.} Alternative view of ranking of the NA methods.  Each
pie chart shows the fraction of evaluation test ranks that fall into
the 1--4, 5--8, and 9--12 rank bins out of all evaluation test ranks
in the given alignment category. For example, for the PE framework, in
the PE-P-P alignment category, 56\%, 26\%, and 18\% of the evaluation
test ranks fall into ranks 1--4, 5--8, and 9--12, respectively,
totaling to 100\% of the evaluation test ranks in the PE-P-P alignment
category.  The pie charts allow us to compare the three alignment
categories rather than individual NA methods in each category. The
larger the pie chart for the better (lower) ranks, and the smaller the
pie chart for the worse (higher) ranks, the better the alignment
category. For example, in the PE framework, PE-P-P has the most
evaluation tests ranked 1--4 and the fewest evaluation tests ranked
9--12, followed by PE-M-P, followed by PE-M-M. This implies that
PE-P-P is superior to PE-M-P and PE-M-M.  The pie charts are color
coded with respect to alignments of network pairs/sets with known and
unknown node mapping, and FQ and TQ measures.  {\bf View III.}
Overall ranking of an NA method versus its running time. The latter
are running time results when aligning all network pairs in the
Y2H$_1$ network set under the PE framework, and when aligning the
Y2H$_1$ network set under the ME framework, where each method is
restricted to use a maximum of 64 cores. The size of each point
visualizes the overall ranking of the corresponding method over all
evaluation tests over all network pairs/sets, corresponding to the
``Overall rank'' column in View I; the larger the point size, the
better the method. In order to allow for easier comparison between the
different alignment categories, ``Average'' shows the average running
times and average rankings of the methods in each alignment
category.}
\label{fig:res_t}
\end{figure*}

\begin{figure}[h!]
\centering
\includegraphics[width=\linewidth]{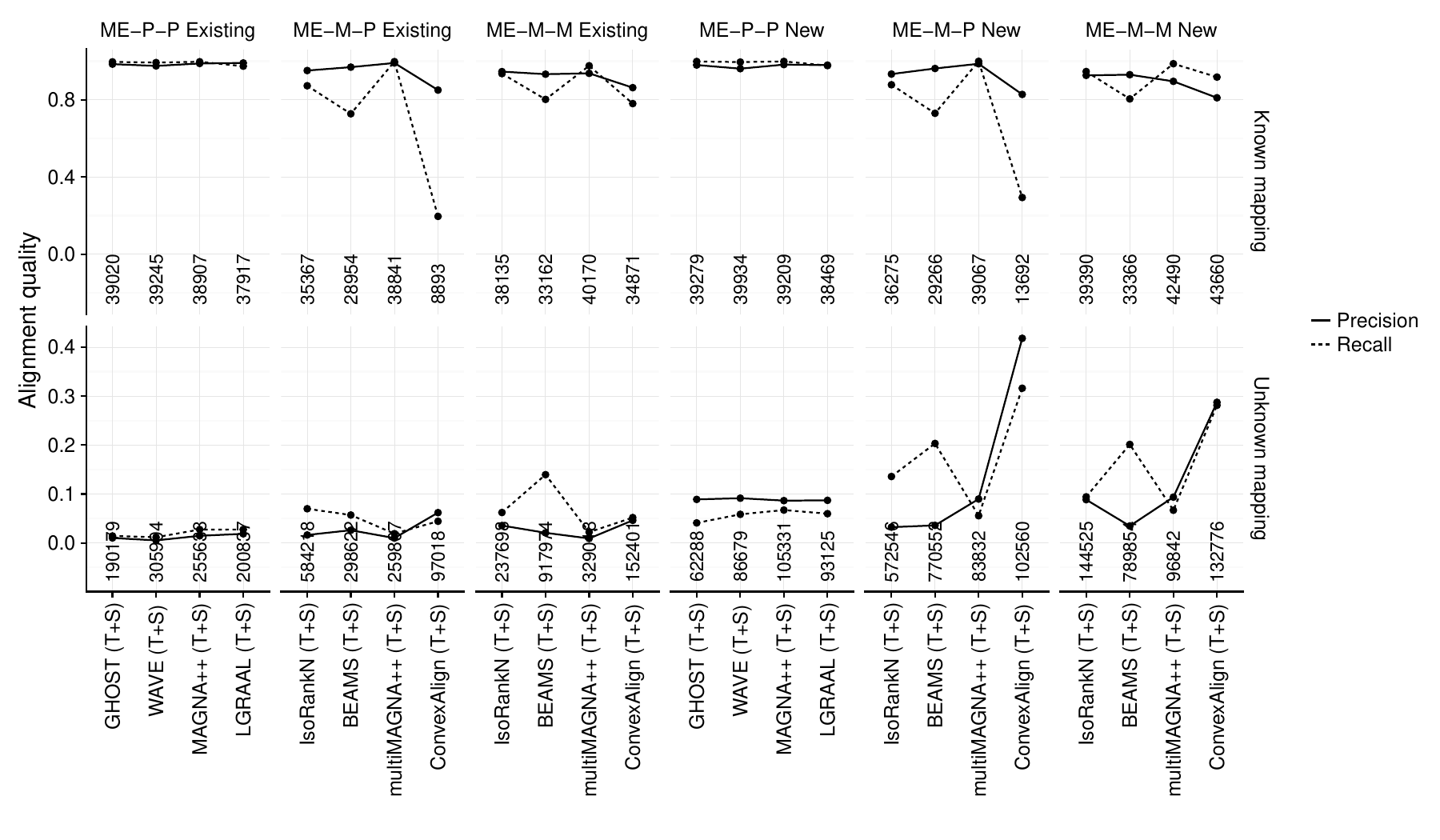}
\caption{Comparison of protein function prediction accuracy between
the {\bf new} (approach 3) versus the
{\bf existing} prediction approach for multiple alignments (approach 2), for all
alignments from the ME framework (i.e., ME-P-P, ME-M-P, and ME-M-M categories).  We calculate the prediction
accuracy as described in Fig. \ref{fig:exp3_oldvsnew} in the main paper. Each column
shows the precision and recall achieved by the new or existing
prediction approach for each NA method, as well as the number of
predictions made by the approach.  The alignments are separated
into networks sets with known and unknown mapping.}
\label{fig:exp3_oldvsnew_full}
\end{figure}

\begin{figure}[h!]
\centering
\includegraphics[width=\linewidth]{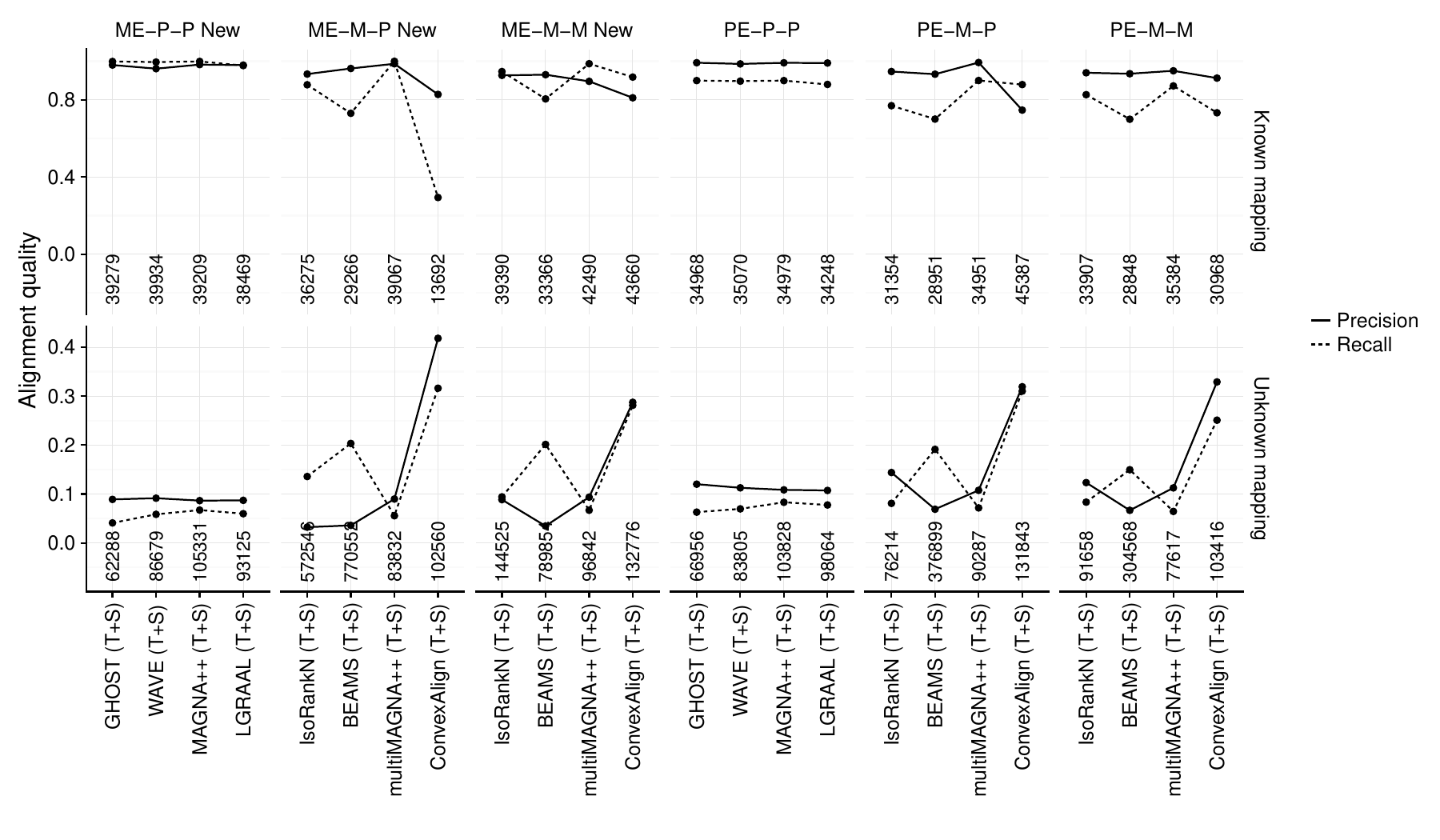}
\caption{Comparison of protein function prediction accuracy under the
{\bf PE framework} (i.e., PE-P-P, PE-M-P, and PE-M-M categories) and {\bf ME
framework} (i.e., ME-P-P, ME-M-P, and ME-M-M categories). We calculate
the prediction accuracy as described in Fig. \ref{fig:exp3_oldvsnew}
in the main paper. Each column shows the precision and recall achieved
by the new or existing prediction approach for each NA method, as well
as the number of predictions made by the approach.  The alignments are
separated into networks sets with known and unknown mapping.}
\label{fig:exp3_pefvsmef_full}
\end{figure}

\begin{figure*}
\centering
(a)\includegraphics[width=0.5\linewidth]{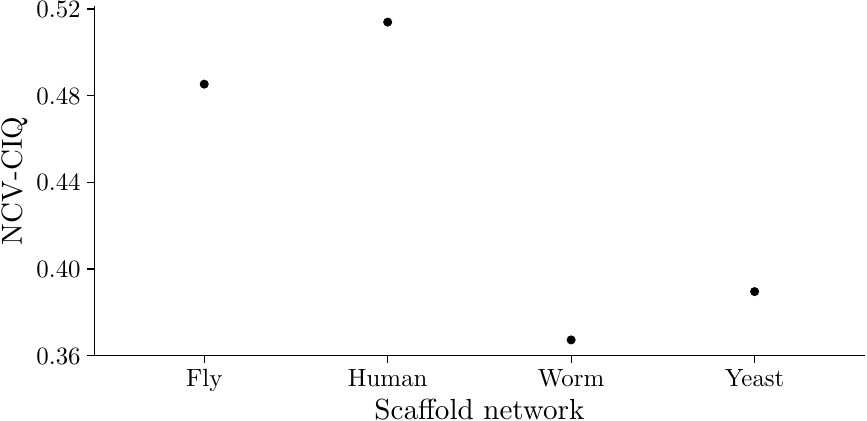}
(b)\includegraphics[width=0.5\linewidth]{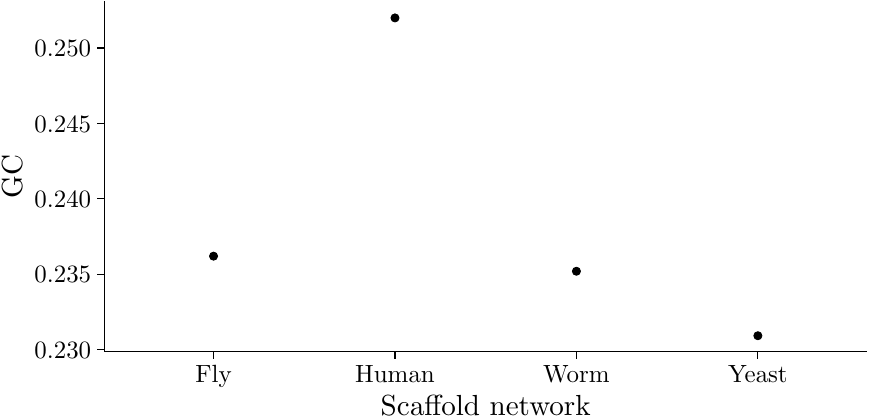}
\caption{
Illustration of the effect of the choice of scaffold network on
alignment quality when combining pairwise alignments into a multiple
alignment. These are representative results for one of the analyzed TQ
measures (NCV-CIQ; panel (a)), one of the analyzed FQ measures (GO
correctness -- GC; panel (b)), one of the analyzed network sets
(Y2H1), and one of the analyzed NA methods (WAVE). Clearly, different 
choices of scaffold network ($x$-axis) yield different alignment
quality scores ($y$-axis). The same holds for other combinations of
alignment quality measures, network sets, and NA methods. In our
evaluation, of all scaffold network choices, the one that yields the
best multiple alignment is chosen. In this particular representative
scenario, it is the human network that was chosen as the scaffold,
since this scaffold choice clearly yields significantly better
alignment quality than any other scaffold choice.  } 
\label{fig:scaf}
\end{figure*}

\begin{figure}[h!]
\centering
\includegraphics[width=0.7\linewidth]{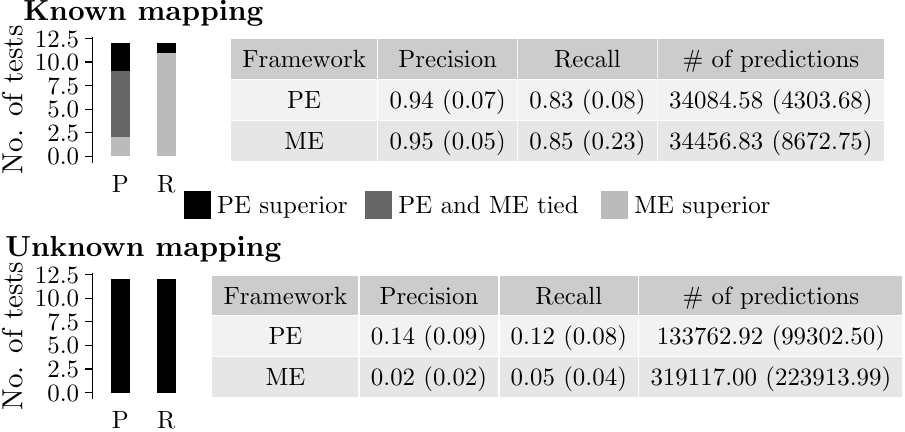}
\vspace{-0.4cm}
\caption{\textcolor{black}{Comparison of protein function prediction accuracy under the
the PE and ME frameworks, where we use approach 2 for the ME framework (rather than using approach 3 for the ME framework like we do in Fig. \ref{fig:exp3_pefvsmef} of the main paper).  The figure can be interpreted the same way
as Fig. \ref{fig:exp3_oldvsnew} in the main paper.}}
\label{fig:exp3_pefvsmef_approach2}
\vspace{-0.4cm}
\end{figure}

%


\clearpage
\newpage

\section*{Funding}

This work was supported by the Air Force Office of Scientific Research
(AFOSR) YIP FA9550-16-1-0147 grant.
\bibliographystyle{apalike}

\bibitem[Alkan and Erten, 2014]{BEAMS}
Alkan, F. and Erten, C. (2014).
\newblock {BEAMS: backbone extraction and merge strategy for the global
  many-to-many alignment of multiple PPI networks}.
\newblock {\em Bioinformatics}, 30(4):531--539.

\bibitem[Dohrmann et~al., 2015]{SMAL}
Dohrmann, J., Puchin, J., and Singh, R. (2015).
\newblock Global multiple protein-protein interaction network alignment by
  combining pairwise network alignments.
\newblock {\em BMC Bioinformatics}, 16(Suppl 13):S11.

\bibitem[Everitt et~al., 2001]{ClusterAnalysis}
Everitt, B.~S., Landau, S., and Leese, M. (2001).
\newblock {\em Cluster {A}nalysis}.
\newblock Wiley.

\bibitem[Gligorijevi\'{c} et~al., 2015]{FUSE}
Gligorijevi\'{c}, V., Malod-Dognin, N., and Pr\v{z}ulj, N. (2015).
\newblock {FUSE: Multiple Network Alignment via Data Fusion}.
\newblock {\em Bioinformatics}, 32(8):1195--1203.

\bibitem[Kuchaiev et~al., 2010]{GRAAL}
Kuchaiev, O., Milenkovi\'{c}, T., Memi\v{s}evi\'{c}, V., Hayes, W., and
  Pr\v{z}ulj, N. (2010).
\newblock {Topological network alignment uncovers biological function and
  phylogeny}.
\newblock {\em Journal of The Royal Society Interface}, 7(50):1341--1354.

\bibitem[Liao et~al., 2009]{IsoRankN}
Liao, C., Lu, K., Baym, M., Singh, R., and Berger, B. (2009).
\newblock Iso{R}ank{N}: {S}pectral methods for global alignment of multiple
  protein networks.
\newblock {\em Bioinformatics}, 25(12):i253--258.

\bibitem[Meng et~al., 2016b]{LocalVsGlobal}
Meng, L., Striegel, A., and Milenkovi\'{c}, T. (2016b).
\newblock Local versus global biological network alignment.
\newblock {\em Bioinformatics}, 32(20):3155--3164.

\bibitem[Milenkovi\'{c} and Pr\v{z}ulj, 2008]{Milenkovic2008}
Milenkovi\'{c}, T. and Pr\v{z}ulj, N. (2008).
\newblock Uncovering biological network function via graphlet degree
  signatures.
\newblock {\em Cancer Informatics}, 6:257--273.

\bibitem[Pr\v{z}ulj, 2007]{GDD}
Pr\v{z}ulj, N. (2007).
\newblock Biological network comparison using graphlet degree distribution.
\newblock {\em Bioinformatics}, 23(2):e177--e183.

\bibitem[Saraph and Milenkovi\'{c}, 2014]{MAGNA}
Saraph, V. and Milenkovi\'{c}, T. (2014).
\newblock {MAGNA}: Maximizing accuracy in global network alignment.
\newblock {\em Bioinformatics}, 30(20):2931--2940.

\bibitem[Vijayan and Milenkovi\'{c}, 2016]{multiMAGNA++}
Vijayan, V. and Milenkovi\'{c}, T. (2016).
\newblock Multiple network alignment via {multiMAGNA++}.
\newblock In {\em Proc. of Workshop on Data Mining in Bioinformatics (BIOKDD)
  at the Conference on Knowledge Discovery and Data Mining (KDD)}.

\bibitem[Vinh et~al., 2007]{PartitionSimilarity}
Vinh, N.~X., Epps, J., and Bailey, J. (2007).
\newblock {Information theoretic measures for clusterings comparison: Variants,
  properties, normalization and correction for chance}.
\newblock {\em The {J}ournal of {M}achine {L}earning {R}esearch}, 11:410--420.

\end{document}